\documentclass[aps,prd,showpacs,nofootinbib]{revtex4-1}
\usepackage{latexsym}
\usepackage{stmaryrd}
\usepackage{amsmath,amsfonts}
\usepackage{amsbsy}
\usepackage{mathrsfs}
\usepackage{color}
\usepackage{psfrag}
\usepackage{enumerate}

\usepackage{amsmath,amssymb,calc,amsfonts}
\usepackage{latexsym}

\usepackage{graphicx,calc,epsfig}
\def\ut#1{\rlap{\lower1ex\hbox{$\sim$}}#1{}}
\newcommand{\N}{\mathbb{N}}

\newcommand{\R}{\mathbb{R}}
\newcommand{\n}{\nonumber}
\newcommand{\be}{\nopagebreak[3]\begin{equation}}
\newcommand{\ee}{\end{equation}}
\newcommand{\ba}{\nopagebreak[3]\begin{eqnarray}}
\newcommand{\ea}{\end{eqnarray}}
\DeclareFontFamily{U}{rsfs}{}         
\DeclareFontShape{U}{rsfs}{m}{n}{<5> rsfs5 <6><7> rsfs7          %
  <8><9><10><10.95><12><14.4><17.28><20.74><24.88> rsfs10}{}     %
\DeclareMathAlphabet{\mathfs}{U}{rsfs}{m}{n}                     %
\newcommand{\mfs}[1]{\mathfs {#1}}                               %
\newcommand{\inter}{{\lrcorner}}
\newcommand{\va}{\scriptscriptstyle}
\newcommand{\van}{\scriptstyle}

\newcommand{\sH}{{\mfs H}}
\newcommand{\sL}{{\mfs L}}

\newcommand{\sM}{{\mfs M}}

\newcommand{\sI}{{\mfs I}}
\newcommand{\sD}{{\mfs D}}\newcommand{\sO}{{\mfs O}}

\newcommand{\Lie}{\sL}

\newcommand{\Vect}{\mathrm{Vect}}

\def\i{i}

\def\pb#1{\rlap{\lower1.5ex\hbox{$\longleftarrow$}}{#1}}
\def\dpb#1{\rlap{\lower1.5ex\hbox{$\Longleftarrow$}}{#1}}
\def\spb#1{\rlap{\lower1.5ex\hbox{$\leftarrow$}}{#1}}
\def\sdpb#1{\rlap{\lower1.5ex\hbox{$\Leftarrow$}}{#1}}

\begin{document}

\title{ Static isolated horizons: $SU(2)$ invariant phase space, quantization, and black hole entropy}

\date{\today}

\author{Alejandro Perez, Daniele Pranzetti }

\affiliation{Centre de Physique Th\'eorique\footnote{Unit\'e
Mixte de Recherche (UMR 6207) du CNRS et des Universit\'es
Aix-Marseille I, Aix-Marseille II, et du Sud Toulon-Var; laboratoire
afili\'e \`a la FRUMAM (FR 2291)}, Campus de Luminy, 13288
Marseille, France.}

\begin{abstract}
We study the classical field theoretical formulation of static generic isolated horizons in a manifestly $SU(2)$ invariant formulation. We 
show that the usual classical description requires revision in the non-static case  due to the breaking of diffeomorphism invariance at the horizon 
leading to the non conservation of the usual 
pre-symplectic structure. We argue how this difficulty could be avoided by a simple enlargement of the field content at the horizon that restores diffeomorphism invariance.
Restricting our attention to static isolated horizons we study the effective theories describing the boundary  degrees of freedom.
A quantization of the horizon degrees of freedom is proposed. By defining a statistical mechanical ensemble where only the area 
$a_{\va H}$ of the horizon is fixed macroscopically---states with fluctuations away from spherical symmetry are allowed---we show that it is possible to obtain agreement with 
the Hawkings area law ($S=a_{H}/(4\ell_p^2)$) without fixing the Immirzi parameter to any particular value: consistency with the area law only imposes a relationship between 
the Immirzi parameter and the level of the Chern-Simons theory involved in the effective description of the horizon degrees of freedom. 
\end{abstract}

\maketitle

\section{Introduction}

One of the most striking predictions of General Relativity (GR) is
the creation of Black Holes (BH) as the final stage of gravitational
collapse. A prediction that is by now supported by an important
accumulation of observational evidence
\cite{observ1,observ2,observ3}. However, despite the extremely
simple general relativistic description of the spacetime geometry of
the black hole (BH) external region---at least in the idealized
scenario given by the Kerr-Newman solutions describing the situation
once the dynamical phase of collapse has settled down---the power of
general relativity reduces drastically when it comes to the
understanding of the internal dynamics, as singularities of the
spacetime geometry are unavoidably developed. Simple arguments imply
that a complete consistent description of the gravitational collapse
would necessarily require dealing with quantum fluctuations of the
gravitational interaction described by some quantum theory of
gravity.

However, not only the interior BH physics---unaccessible to external
observers---is calling for a theory of quantum gravity (QG). As
shown by the celebrated works of Bekenstein and Hawking (\cite
{Bekenstein, Hawking}), there are strong theoretical arguments
indicating that idealized BHs in their stationary phase behave as
thermodynamical systems, with their own analogue of the {\it
zeroth}, {\it first}, {\it second}, and {\it third law} of
thermodynamics, respectively. More precisely one has that:
{\em(0)}~The surface gravity $\kappa_{\va H}$ on the event horizon
of stationary BH is constant. {\em(1)}~Under external perturbation
the initially stationary state of a black hole can change but the
final stationary state will be described by another Kerr-Newman
solution whose parameters (mass $M$, electric charge $Q$ and angular
momentum $J$) readjust according to
      \be
      \delta M = \frac{\kappa_{\va H}}{8\pi G} \delta a_{\va H}+\Phi_{\va H}\delta Q +\Omega_{\va H}\delta J\,
      \ee
where $a_{\va H}$ is the horizon area,  $\Phi_{\va H}$ is the
electrostatic potential at the horizon, and $\Omega_{\va H}$ the
angular velocity of the horizon. {\em(2)}~The BH horizon area can
only increase
      \be
      \delta a_{\va H}\geq 0\,
      \ee
{\em(3)}~No finite physical process can allow a BH to become
extremal ($\kappa_{\va H}=0$).

This analogy with thermodynamics became strict with the discovery of
the Hawking effect which implies that stationary BHs radiate as
black bodies with a temperature $T= \kappa_{\va H}/(2\pi)$ and
hence, through the first law, have an entropy $S=a_{\va
H}/(4\ell_p^2)$. Understanding the physical nature of the
microscopic degrees of freedom leading to such entropy requires a
quantum description of the gravitational field. In this paper we
study the problem from the perspective of loop quantum gravity
\cite{lqg1,lqg2,lqg3,lqg4}.

However, the very notion of black hole---as the region causally
disconnected from future null infinity---becomes elusive in the
context of quantum gravity due to the simple fact that black hole
radiation in the semiclassical regime imply that in the full quantum
theory the global structure of space-time (expected to make sense
away from the strong field region) might completely change. In fact,
recent models in two dimensions support the view that this is the
case \cite{abhay2d}. For that reason, the problem of black hole
entropy in quantum gravity requires the use of a local or
quasi-local notion of horizon in equilibrium.

In recent years such a local definition of BH has been introduced
(\cite{ACK}) through the concept of Isolated Horizons (IH). Isolated
horizons are regarded as a sector of the phase-space of GR
containing a horizon in ``equilibrium'' with the external matter and
gravitational degrees of freedom. This local definition has been
first used for the black-hole entropy calculation (for spherically
symmetric IH) in \cite{Ashtekar-Baez-Krasnov} in the context of loop
quantum gravity. In this seminal work the authors show, after
introduction of a suitable gauge fixing, how the degrees of freedom
that are relevant for the entropy calculation can be encoded in a
boundary $U(1)$ Chern-Simons theory. Based on this work, state
counting of horizon states leads to agreement with the
Bekenstein-Hawking formula upon appropriate tuning of the Immirzi parameter
\cite{counting01,counting02,counting03,counting04,counting11,counting12,counting13,counting14}
(for a complete description of the beautiful counting techniques
used in these calculations see~\cite{counting3}).

A more simple, natural and effective description of the boundary
degrees of freedom for spherically symmetric IH can be obtained in
terms of an $SU(2)$ Chern-Simons theory \cite{prl, Spherically}.
This latter treatment clarifies how the $U(1)$ gauge reduction
previously used leads to the over counting of allowed horizon states
\cite{prl}. As a consequence the entropy calculation changes:
agreement with the area law requires the tuning of the Immirzi
parameter to a different value, and the universal logarithmic
corrections change from $\Delta S_{\va U(1)}=-1/2 \log a_{\va H}$ to
$\Delta S_{\va SU(2)}=-{3/2} \log a_{\va H}$ now in agreement with
other approaches
\cite{other11,other12,other13,other21,other22,other23,other24,Terno}.
However, there remains the somewhat unnatural spherical symmetry
requirement at the classical level that one would like to eliminate.

In this work we extend the $SU(2)$ invariant treatment to a wider
class of IH containing distortion. More precisely we consider
generic isolated horizons \cite{gih1,gih2} of the {\em static} type
and show how they too admit an effective description in terms of
$SU(2)$ Chern-Simons theories similar in spirit to what was found in
the spherically symmetric case. Our approach is fundamentally
different from the one of \cite{AEV} where---thanks to the
additional assumption of axisymmetry (not necessary in the present
treatment)---the system is mapped to a model equivalent to the Type
I case if the multipole moments describing the amount of distortion
are fixed classically. Recently the treatment of \cite{AEV} has been
generalized to generic isolated horizons in \cite{jony} in a way
that allows to remove all symmetry assumptions, where the main idea
remains to  describe the boundary degrees of freedom in terms of a
canonical connection (called area-connection in the second
reference).
In the present treatment, no symmetry assumption is necessary either
(Type I, Type II, and Type III horizons are all treated on equal
footing), only staticity is a necessary condition for the dynamical
system to be well defined (see Section \ref{sec:conserved symplectic
structure}).  Our approach is different from previous work dealing
with distorted IH \cite{AEV,jony} in two main respects: first the
treatment is $SU(2)$ gauge invariant avoiding in this way the
difficulties found upon quantization in the gauge fixed $U(1)$
formulation, and second, distortion is not erased by the choice of a
mapping to a canonical Type I connection. In particular, the degrees
of freedom related to distortion are encoded in observables of our
system and can be quantized. In this new treatment we can find the
old Type I theory in the sense that when we define the statistical
mechanical ensemble by fixing the macroscopic area $a_{\va H}$ and
imposing spherical symmetry, we get an entropy consistent with the
one in \cite{Spherically}. Moreover, we can go beyond this
observation computing the entropy for an ensemble where only the
area is fixed macroscopically while distortion is allowed to
fluctuate (see also \cite{jony}).

We will show that a one-parameter ambiguity arises when one
describes the classical boundary theory in terms of $SU(2)$
connections. This ambiguity is analogous to the appearance of the
Immirzi parameter in the description of the phase space of general
relativity in terms of $SU(2)$ connection variables. More precisely,
the horizon degrees of freedom will be described by a pair of
$SU(2)$ Chern-Simons theories with the same level $k$ which is
otherwise arbitrary. Hence, the classical ambiguity  referred to
above is encoded in the value of the level $k\in \N$ of the
Chern-Simons theories. We show that one can recover the
Bekenstein-Hawking entropy without the need of fixing the Immirzi
parameter to a particular value. Instead, (the semiclassical)
Hawking's area law `dictates' the relationship between the Immirzi
parameter in the bulk theory (LQG)  and the analog in the  boundary
theory (the Chern-Simons level), in the sense that the Immirzi
parameter can now take different discrete values according to the
choice of the level $k\in \N$.

The paper is organized as follows: In Section \ref{defini} we
restate the definition of isolated horizons as put in
\cite{Spherically} and based on \cite{gih1,gih2}, and include a new
classification of isolated horizons according to the reality of the
Weyl tensor component $\Psi_2$ (this classification is already
implicit in \cite{gih1,gih2}). In Section~\ref{ke} we introduce the
basic equations that follow from the definition of isolated
horizons. In Section \ref{sec:conserved symplectic structure} we
construct the pre-symplectic structure of static isolated horizons
and prove its conservation. We also show that the usual treatment
used so far in the literature on isolated horizons cannot be
directly applied in the case of non-static horizons and requires a
modification. We propose a possible way to deal with this problem;
however, a full analysis of this issue is outside the scope of this
paper. In Section \ref{Quantization} we quantize the system and show
that the subset of spherically symmetric states in the Hilbert space
are in one-to-one correspondence with the admissible states in the
formulation of reference \cite{Spherically}. Thus, if spherical
symmetry is imposed as a condition defining the ensemble of states
in the statistical mechanical treatment, the entropy comes out
proportional to $a_{\va H}$ as expected. In Section \ref{ec} we
compute the entropy starting from an ensemble where only the area is
fixed macroscopically and distortion is allowed, we show that
compatibility with the area law can be obtained and, if an
appropriate and natural paradigm shift is undertaken, there is no
need to fix the Immirzi parameter to a given specific value
(Subsection~\ref{ps}). Concluding remarks are presented in Section
\ref{conclusion}.


\section{Definition of Isolated Horizons}\label{defini}

There are some parts of this section that literally follow reference
\cite{Spherically}.

The standard definition of a BH as a spacetime region from which no
information can reach idealized observers at (future null) infinity
is a global definition. This notion of BH requires a complete
knowledge of a spacetime geometry and is therefore not suitable for
describing local physics.  On physical grounds a quasilocal
definition is necessary for quantum considerations. That this should
be the case is clear from the fact that physical black holes are
expected to radiate due to the Hawking effect and hence the usual
mathematical definition might not even make sense. The first
quasilocal definitions in loop quantum gravity  were introduced in
\cite{ACK,gih1,gih2} with the name of isolated horizons (IH). Here
we present this definition according to
\cite{gih1,gih2,afk,abl2001}.  A full discussion of the geometrical
meaning of the following conditions
can be found in \cite{gih1,gih2, Spherically}.\\

\textit{Definition:} The internal boundary $\Delta$ of a history $(\sM, g_{ab})$ will be
called an \textit{isolated horizon} provided the following conditions hold:
\begin{enumerate}[i)]
\item \textit{Manifold conditions:} $\Delta$ is topologically $S^2 \times R$, foliated by a (preferred)
family of 2-spheres $S$ and equipped with an equivalence class $[\ell^a]$ of transversal
future pointing null vector fields whose flow preserves the foliation, where $\ell^a$ is equivalent to 
$\ell'^a$ if $\ell^a = c \ell'^a$ for some positive real number $c$.

\item  \textit{Dynamical conditions:} All field equations hold at $\Delta$.

\item  \textit{Matter conditions:}
On $\Delta$ the stress-energy tensor $T_{ab}$ of matter is such that
$-T^a{}_b\ell^b$ is causal and future directed.

\item  \textit{Conditions on the metric $g$ determined by $e$, and on its levi-Civita derivative
operator $\nabla$:} (iv.a) The expansion of $\ell^a$ within $\Delta$ is zero.
%
%
%
This, together with the energy condition (iii) and the Raychaudhuri equation at $\Delta$, ensures
that $\ell^a$ is additionally shear-free.  This in turn implies that the Levi-Civita derivative
operator $\nabla$ naturally determines a derivative operator $D_a$ intrinsic to $\Delta$ via
$X^a D_a Y^b := X^a \nabla_a Y^b$, $X^a, Y^a$ tangent to $\Delta$. We then impose (iv.b) $[\Lie_\ell, D] = 0$.
%
%

\item  \textit{Restriction to `good cuts.'}
One can show furthermore that $D_a \ell^b = \omega_a \ell^b$ for some $\omega_a$ intrinsic to $\Delta$.
A 2-sphere cross-section $S$ of $\Delta$ is called a `good cut' if the pull-back of $\omega_a$ to $S$ is
divergence free with respect to the pull-back of $g_{ab}$ to $S$.  As shown in \cite{gih1}, 
every horizon
satisfying (i)-(iv) above possesses at least one foliation into `good cuts'; this foliation is furthermore
generically unique.  We require that the fixed foliation coincide with a foliation into `good cuts.'

\end{enumerate}

\subsection{Isolated horizon classification according to their symmetry groups}

Next, let us examine symmetry groups of isolated horizons. A \textit{symmetry} of
$(\Delta, q, D, [\ell^a])$ is a diffeomorphism on $\Delta$ which preserves the horizon geometry
$(q, D)$ and at most rescales elements of $[\ell^a]$ by a positive constant. It is clear that 
diffeomorphisms generated by $\ell^a$ are symmetries. So, the symmetry group $G_\Delta$ is at 
least 1-dimensional. In fact, there are only three possibilities for $G_\Delta$:\\
\begin{enumerate}[(a)]
\item Type I: the isolated horizon geometry is spherical; in this case, $G_\Delta$ is four dimensional ($SO(3)$ rotations plus rescaling-translations\footnote{In a coordinate system where $\ell^a=(\partial/\partial v)^a$ the rescaling-translation corresponds to the affine map $v\to c v+b$ with $c,b \in \R$ constants. } along $\ell$);\\
\item Type II: the isolated horizon geometry is axi-symmetric; in this case, $G_\Delta$ is two dimensional
(rotations round symmetry axis plus rescaling-translations along $\ell$);\\
\item Type III: the diffeomorphisms generated by $\ell^a$ are the only symmetries; $G_\Delta$ is one dimensional.
\end{enumerate}

Note that these symmetries refer only to the horizon geometry. The
full space-time metric need not admit any isometries even in a
neighborhood of the horizon. 

\subsection{Isolated Horizons Classification According to the Reality of $\Psi_2$}

\begin{enumerate}[(a)]
\item Static: In the Newman-Penrose formalism
(in the null tetrads adapted to the IH geometry introduced in the
following section) static isolated horizons are characterized by the
condition\be{\rm Im}(\Psi_2)=0\label{stati}\ee on the Weyl tensor
component $\Psi_2=C_{abcd} \ell^a m^b \bar m^c n^d$. One can then
show that for this class of isolated horizons (see next section) the
pull-back to $H$ (the preferred family of sections) of $v\inter
K_i\Sigma^i$ vanishes for all $v\in T(H)$. This corresponds to
having the horizon locally ``at rest''. In the axisymmetric case,
according to the definition of multiple moments of Type II horizons
constructed in \cite{AEV},  {\em static isolated horizons} are {\em
non-rotating} isolated horizons, {\em i.e.}, those for which all
angular momentum multiple moments vanish. Static black holes (e.g.,
those in the Reissner-Nordtrom family) have static isolated
horizons. There are Type I, II and III static isolated horizons.

\item  Non-Static: In the Newman-Penrose formalism (in the null tetrads
adapted to the IH geometry introduced in the following section)
non-static isolated horizons are characterized by the
condition\be{\rm Im}(\Psi_2)\not=0\ee The pull-back to $H$ (the
preferred family of sections) of $v\inter K_i\Sigma^i$ does not
vanish for all $v\in T(H)$. The horizon is locally ``in motion''.
The Kerr black hole is an example of this type.
\end{enumerate}

In this paper we will construct the conserved pre-symplectic
structure of static isolated horizons  (no symmetry assumptions on
the horizon are made). We will also show that the usual
pre-symplectic structure is not conserved in the presence of a
non-static black hole, which implies that a complete treatment of
non-static isolated horizons (including rotating isolated horizons)
remains open. In this direction, we propose some general ideas
leading to a conserved symplectic structure for non-static isolated
horizons and the restoration of diffeomorphism invariance. However,
due to the very different nature of such approach, the quantization
of such proposals is left for future studies.

\section{Some Key Equations}\label{ke}
In this section we use the definition of isolated horizons provided
in the previous section to prove some of the equations we will need
in the sequel. General relativity  in the first order formalism is
described in terms a tetrad of four 1-forms $e^I$ ($I=0,3$ internal
indices) and a Lorentz connection $\omega^{IJ}=-\omega^{JI}$. The
metric can be recovered by \be g_{ab}=e^I_ae^J_b\eta_{IJ}\, \ee
where $\eta_{IJ}=diag(-1,1,1,1)$.
In the time gauge, where the tetrad $e^I$ is such that $e^0$ is a
time-like vector field normal to $M$, the three $1$-forms
$K^i=\omega^{0i}$ play a special role in the parametrization of the
phase space. In particular the so-called Ashtekar connection is \be
A^{+i}_a=\Gamma^i_a+\i K^i_a \ee where $\Gamma^i$ is the spin
connection satisfying Cartan's first equation $d_\Gamma e^i=0$. We
also introduce\begin{equation}\nonumber \Sigma^{IJ}\equiv e^I\wedge
e^J\, \ \ {\rm and} \ \ \ \Sigma^{+i}\equiv
\epsilon^i\,_{jk}\Sigma^{jk}+2 i \Sigma^{0i}
\end{equation}
and $F^i(A)$ the curvature of the connection $A^i$. At $H$, we will
also often work in the gauge where $e^1$ is normal to $H$ and $e^2$
and $e^3$ are tangent to $H$. This choice is only made for
convenience, as the equations are all gauge covariant, their
validity in one frame implies their validity in all frames.

\vspace{12pt} \noindent {\bf Statement 1:}  In the gauge where the
tetrad is chosen so that  $\ell^a=2^{-1/2} (e^a_0+e^a_1)$ (which can
be completed to a null tetrad $n^a=2^{-1/2} (e^a_0-e^a_1)$, and
$m^a=2^{-1/2} (e^a_2+i e^a_3)$),  the  shear-free and vanishing
expansion (condition ($iv.a$) in the definition of IH)  imply \be
\sdpb{\omega}^{21}=\sdpb{\omega}^{20} \ \ \ {\rm and}\ \ \
\sdpb{\omega}^{31}=\sdpb{\omega}^{30} \label{l1}\ee \noindent {where
the double arrow means ``pull-back to the horizon''.} The proof of
this statement
 can be found in~\cite{Spherically}.

\vspace{12pt} \noindent  {\bf Statement 2:} We start from the
identity (that can be derived from Cartan's second structure
equations) \be F_{ab}{^i}(A^{\va +})=-\frac{1}{4} R_{ab}^{\ \ cd}
\Sigma^{\va + i}_{cd} \ee where $R_{abcd}$ is the Riemann tensor and
$\Sigma^{\va + i}= \epsilon^{i}_{\ jk} e^j\wedge e^k+ i2 e^0\wedge
e^i$. A simple algebraic calculation using the null tetrad formalism
(see for instance \cite{Chandra} page 43) with the null tetrad of
Statement 1, and the definitions $\Psi_2=C_{abcd} \ell^a m^b \bar
m^c n^d$ and $\Phi_{11}=R_{ab}(\ell^an^b+m^a\bar{m}^b)/4$, where
$R_{ab}$ is the Ricci tensor and $C_{abcd}$ the Weyl tensor, yields
 \be
 \sdpb{F_{ab}}^i(A^{\va +})=(\Psi_2-\Phi_{11}-\frac{R}{24})\sdpb{\Sigma}^i_{ab}
 \ee
where, on $M$, $\Sigma^{i}={\rm Re}[\Sigma^{{\va +}i}]=
\epsilon^{i}_{\ jk} e^j\wedge e^k $. For simplicity, here we will
assume that no matter is present at the horizon, $\Phi_{11}=R=0$,
hence \be\label{UNO}
 \sdpb{F_{ab}}^i(A^{\va +})=\Psi_2\ \sdpb{\Sigma}^i_{ab}
 \ee
An important point here is that the previous expression is valid for
any two sphere $S^2$ (not necessarily a horizon) embedded in
spacetime in an adapted null tetrad where $\ell^a$ and $n^a$ are
normal to $S^2$.

{ Here we will concentrate, for simplicity, in the pure gravity
case. In this special case, and due to the vanishing of both the
expansion and shear of the generators congruence $\ell^a$, the Weyl
component $\Psi_2$ at the Horizon is simply related to the Gauss
scalar curvature $R^{\va (2)}$of the  two spheres

 {

\vspace{12pt} \noindent  {\bf Statement 3:} The statement 1 has an
immediate consequence for static isolated horizons: the reality of
$\Psi_2$ implies, for the component $i=1$ in the frame of the
statement 1,  that $\sdpb{dK}^1=0$. The good-cut condition ($v$) in
the definition then implies that \be \sdpb{K}^1=0\label{k1} \ee

\vspace{6pt} \noindent  {\bf Statement 4:}  For static isolated
horizons we have that \be\label{eq:KK=lambda_0 Sigma}
\sdpb{K^j}\wedge \sdpb{K^k}\epsilon_{ijk}=c \, \sdpb{\Sigma}^i\, \ee
 for $c: H\rightarrow \R$ an extrinsic curvature scalar.

\vspace{12pt} \noindent {\em Proof:} In order to simplify the
notation all free indices associated to forms that appear in this
proof are pulled back to $H$ (this allows us to drop the double
arrows from equations). In the frame of statement 1,  where  $e^1$
is normal to $H$,  the only non-trivial component of the  equation
we want to prove  is the $i=1$ component, namely:
 \be
{K^A}\wedge {K^B}\epsilon_{AB}=c\, {\Sigma}^1  \label{l2}\ee where
$A,B=2,3$ and $\epsilon^{AB}=\epsilon^{1AB}$. Now,  in that gauge,
we have that $K^A=c^A_{\ B} e^B$ for some matrix of coefficients
$c^A_{\ B}$. Notice that the left hand side of the previous equation
equals $\det(c) e^A\wedge e^B\epsilon_{AB}$. We only need to prove
that $\det(c)$ is time independent, {\em i.e.}, that
$\ell(\det{c})=0$.
We need to use the isolated horizon boundary condition \be
[\mathscr{L}_\ell, D_b] v^a=0 \ \ \ v^a\in T(\Delta) \ee where $D_a$
is the derivative operator determined on the horizon by the
Levi-Civita derivative operator $\nabla_a$. One important property
of the commutator of two derivative operators is that it also
satisfy the Leibnitz rule (it is itself a new derivative operator).
Therefore, using the fact that  the null vector $n^a$ is normalized
so that $\ell\cdot n=-1$ we get \be 0=[\mathscr{L}_\ell, D_b] \ell^a
n_a=n_a[\mathscr{L}_\ell, D_b] \ell^a +\ell^a [\mathscr{L}_\ell,
D_b] n_a \ \ \ \Rightarrow \ \ \ \ell^a [\mathscr{L}_\ell, D_b]
n_a=0\label{12} \ee 
where we have also used that $\ell^a\in
T(\Delta)$. Now, if one introduces a coordinate $v$ on $\Delta$ such
that $\ell^a \partial_a v = 1$ and $v = 0$ on some leaf of the
foliation, then it follows---from the fact that $\ell$ is a symmetry
of the horizon geometry $(q,D)$, and the fact that the horizon
geometry uniquely determines the foliation into `good cuts'---that
$v$ will be constant on all the leaves of the foliation.  As $n$
must be normal to the leaves one has  $n = - dv$, whence $dn = 0$.
From this it follows that $\mathscr{L}_\ell n=\ell \inter dn+d(\ell
\inter n)=0$ and, therefore, evaluating Equation (\ref{12}) on the
right hand side explicitly, we get
 \ba\n 0&=& \ell^a [\mathscr{L}_\ell, D_b] n_a =\ell^a \mathscr{L}_\ell (D_b n_a)=-\frac{1}{\sqrt{2}}\ell^a \mathscr{L}_\ell (D_b [e^1_a+e^0_a]) \\
 &=& \frac{1}{\sqrt{2}}\ell^a \mathscr{L}_\ell (\omega^1_{b\ \mu} e^{\mu}_a+\omega^{0}_{b\ \mu}e^{\mu}_a)= -\frac{1}{\sqrt{2}}\ell^a \mathscr{L}_\ell (\omega_b^{10} [e^{0}_a+e^{1}_a])+\frac{1}{\sqrt{2}}\ell^a \mathscr{L}_\ell(\omega^1_{b\ A} e^{A}_a+\omega^{0}_{b\ A}e^{A}_a)\n \\
 &=& \ell^a \mathscr{L}_\ell (\omega_b^{10}) n_a \n
\ea where in the second line we have used the fact that
$D_be_a^\nu=-\omega^{\nu}_{b\ \mu} e_a^{\mu}$ plus the fact that as
$\sL_{\ell}q_{ab}=0$ the Lie derivative $\sL_{\ell} e^{A}=\alpha
\epsilon^{AB} e_{B}$ for some $\alpha$ (moreover, one can even fix
$\alpha=0$ if one wanted to by means of internal gauge
transformations). Then it follows that \be \sL_{\ell} K^1=0 \ee a
condition that is also valid for the so called {\em weakly isolated
horizons} \cite{gih1,gih2}. A similar argument as the one given in
Equation (\ref{12})---but now replacing $\ell^a$ by $e^a_{B}\in
T(\Delta)$ for $B=2,3$---leads to
\ba\n 0&=& e_B^a [\mathscr{L}_\ell, D_b] n_a =e_B^a \mathscr{L}_\ell (D_b n_a)=-\frac{1}{\sqrt{2}}e_B^a \mathscr{L}_\ell (D_b [e^1_a+e^0_a]) \\
 &=& \frac{1}{\sqrt{2}}e_B^a \mathscr{L}_\ell (\omega^1_{b\ \mu} e^{\mu}_a+\omega^{0}_{b\ \mu}e^{\mu}_a)= -\frac{1}{\sqrt{2}}e_B^a \mathscr{L}_\ell (\omega_b^{10} [e^{0}_a+e^{1}_a])+\frac{1}{\sqrt{2}}e_B^a \mathscr{L}_\ell(\omega^1_{b\ A} e^{A}_a+\omega^{0}_{b\ A}e^{A}_a)\n \\
 &=& {\sqrt{2}}e_B^a \mathscr{L}_\ell(\omega^{0}_{b\ A}e^{A}_a)={\sqrt{2}} [\mathscr{L}_\ell(\omega_b^{0B}) + \alpha \epsilon^{BA} \omega_b^{0A}]\n
\ea where, in addition to previously used identities, we have made
use of statement 1, Equation (\ref{l1}). The previous equations
imply that the left hand side of Equation (\ref{l2}) is Lie dragged
along the vector field $\ell$, and since $\Sigma^i$ is also Lie
dragged (in this gauge), all this implies that $\sL_{\ell}
(\det(c))=\ell(\det(c))=0$. $\square$

\vspace{12pt} \noindent {\bf Remark 1:} In the GHP formalism
\cite{ghp}, a null tetrad formalism compatible with the choice of
tetrad of statement 1 and hence the IH system, the scalar curvature
of the two-spheres normal to $\ell^a$ and $n^a$ is given~by \be
 R^{\va (2)}=K+\bar K
 \ee where $K=\sigma\sigma'-\rho\rho'-\Psi_2+R+\Phi_{11}$, while  $\sigma$, $\rho$, $\sigma'$ and $\rho'$ denote spin shear and expansion spin coefficients associated with $\ell^a$ and $n^a$ respectively . The shear-free and expansion-free conditions in the definition of IHs translate into $\rho=0=\sigma$ in the GHP formalism}, namely
 \be\label{scalar-curvature}
 R^{\va (2)}=-2 \Psi_2
 \ee
Similarly, the curvature scalar $c$ in (\ref{eq:KK=lambda_0 Sigma})
can be expressed in terms of spin coefficients as \be
c=\frac{1}{2}{(\rho'\bar\rho'-\sigma'\bar\sigma')}\ee which is
invariant under null tetrad transformations fixing $\ell^a$ and
$n^a$.}

\vspace{12pt}
\noindent {\bf Remark 2:} The quantity $d\equiv2\Psi_2+c$ will play a central role in what follows. From the
discussion above we observe that it can be rewritten as \be
d=c-R^{\va(2)}\label{gauss}\ee  {\em i.e.}, it is given by the
difference between Gauss scalar curvature of the horizon---encoding
the local intrinsic geometry of the cross-sections---and the
extrinsic curvature invariant $c$. This quantity will be often
referred to as the {\em distortion} of static isolated horizon in
the rest of this paper. In the entropy calculation we will
considered an statistical mechanical ensemble where only the horizon
area is fixed macroscopically, thus, states with all possible values
of the above local quantity (allowed by the quantum theory) are
counted.

\section{The Conserved Symplectic Structure}\label{sec:conserved symplectic structure}

In this section we prove the conservation of the symplectic
structure of gravity in the presence of an isolated horizon that is
not necessarily spherically symmetric. In this sense, our proof
generalizes the one presented in \cite{Spherically}.

\subsection{The Action Principle}

{ The action principle of general relativity in self dual variables
containing an inner boundary satisfying the IH boundary condition
(for asymptotically flat spacetimes) takes the form \be S[e,A^{\va
+}]=-\frac{i}{\kappa}\int_{\sM} \Sigma^{\va +}_i(e)\wedge F^i(A^{\va
+}) +\frac{i}{\kappa}\int_{\tau_{\infty}} \Sigma^{\va +}_i(e)\wedge
A^{{\va +}i} \label{aacc} \ee where a boundary contribution at a
suitable time cylinder $\tau_{\infty}$ at infinity is required for
the differentiability of the action. {No boundary term is necessary
if one allows variations that fix an isolated horizon geometry up to
diffeomorphisms and Lorentz  transformations \cite{Spherically}.}

First variation of the action yields \be \label{fi}\delta S[e,A^{\va
+}]=\frac{-i}{\kappa}\int_{\sM} \delta \Sigma^{\va +}_i(e)\wedge
F^i(A^{\va +}) -d_{A^{\va +}} \Sigma^{\va +}_{i} \wedge \delta
A^{{\va +}i}+ d( \Sigma^{\va +}_{i} \wedge \delta A^{{\va +}i}
)+\frac{i}{\kappa}\int_{\tau_{\infty}} \delta(\Sigma^{\va
+}_i(e)\wedge A^{{\va +}i}) \ee from which the self dual version of
Einstein's equations follow \ba
\nonumber && \epsilon_{ijk} e^j\wedge F^i(A^{\va +})+ie^0\wedge F_{k}(A^{\va +})=0\\
\nonumber && e_i\wedge F^i(A^{\va +})=0\\
&& d_{A^{\va +}}\Sigma^{\va +}_i=0\label{fe} \ea as the boundary
terms in the variation of the action cancel.}

\subsection{The Conserved Symplectic Structure in Terms of Vector Variables}\label{Conservation}

\begin{figure}[H]
\centerline{\hspace{0.5cm} \(
\begin{array}{c}
\includegraphics[height=4cm]{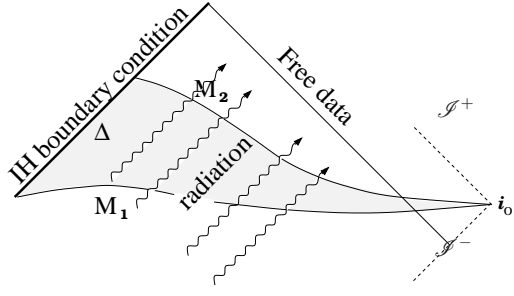}
\end{array}\!\!\!\!\!\!\!\!\!\!\!\! \!\!\!\!\!\!\!\!\!\!\!\! \begin{array}{c}  ^{}\\ \\  \\    \, \sI^{\va +}_{}  \\  \\ \\  \\  \\ \sI^{\va -}\end{array}\) }
\caption{The characteristic data for a (vacuum) spherically symmetric
isolated horizon corresponds to Reissner-Nordstrom data on $\Delta$,
and free radiation data on the transversal null surface with
suitable fall-off conditions. For each mass, charge, and radiation
data in the transverse null surface there is a unique solution of
Einstein-Maxwell equations locally in a portion of the past domain
of dependence of the null surfaces. This defines the phase space of
Type I isolated horizons in Einstein-Maxwell theory. The picture
shows two Cauchy surfaces $M_1$ and $M_2$ ``meeting'' at space-like
infinity $i_0$. A portion of $\sI^+$ and $\sI^-$ are shown; however,
no reference to future time-like infinity $i^+$ is made as the
isolated horizon need not to coincide with the black hole event
horizon. \vspace{-24pt} } \label{figui}
\end{figure}

{ In this work we study general relativity on a spacetime manifold
with an internal boundary satisfying the boundary condition
corresponding to static isolated horizons, and asymptotic flatness
at infinity. The phase space of such system is denoted $\Gamma$ and
is defined by an infinite dimensional manifold where points $p\in
\Gamma$ are given by solutions to Einstein's equations satisfying
the static IH boundary conditions. Explicitly a point $p\in \Gamma$
can be parametrized by a pair $p = ({\Sigma^{\va +}}, {A^{\va +}})$
satisfying the field equations (\ref{fe}) and the requirements of
Definition \ref{defini}. In particular fields at the boundary
satisfy Einstein's equations and the constraints given in Section
\ref{ke}. Let ${\rm T_p} (\Gamma)$ denote the space of variations
$\delta=(\delta\Sigma^{\va +}, \delta A_{\va +})$ at $p$ (in symbols
$\delta\in {\rm T_p} (\Gamma)$).

So far we have defined the covariant phase space as an infinite
dimensional manifold. For it to become a phase space it is necessary
to provide it with a presymplectic structure. As the field
equations, the presymplectic structure can be obtained from the
first variation of the action (\ref{fi}). In particular a symplectic
potential density for gravity can be directly read off from the
total differential term in (\ref{fi}) \cite{cov1,cov2}.} In terms of
the Ashtekar connection and the densitized tetrad, the symplectic
potential density~is \be \theta(\delta)=\frac{-i}{\kappa}\Sigma^{\va
+}_{i} \wedge \delta A^{+i}\ \ \ \ \forall \ \ \delta\in T_p\Gamma
\ee where $\kappa=16\pi G$ and the symplectic current takes the form
\be J(\delta_1,\delta_2)=-\frac{2i}{\kappa}
\delta_{[1}\Sigma^{+}_i\wedge \delta_{2]}A^{ +i}\ \ \ \ \forall \ \
\delta_1, \delta_2 \in T_p\Gamma \ee Einstein's equations imply
$dJ=0$. From Stokes theorem applied to the four dimensional (shaded)
region in Figure \ref{figui} bounded by $M_1$ in the past, $M_2$ in
the future, a timelike cylinder at spacial infinity  on the right,
and the isolated horizon $\Delta$ on the left we
obtain\ba\label{eq:Potential} \delta\mu\equiv-\i\int_{M_1} {\Sigma
}_i\wedge \delta(\Gamma^i+\i K^i)+\i\int_{M_2} {\Sigma }_i\wedge
\delta(\Gamma^i+\i K^i) +\i \int_\Delta {\Sigma}^+_i\wedge \delta
A^{+i} \ea for some functional $\mu$. As we will show now, for
static isolated horizons, the previous equation implies that
symplectic form \be \kappa\,\Omega_{M}(\delta_1,\delta_2)=\int_M
\delta_{[1}\Sigma^i\wedge \delta_{2]} K_i \ee is conserved in the
sense that
$\Omega_{M_2}(\delta_1,\delta_2)=\Omega_{M_1}(\delta_1,\delta_2)$.
In order to prove the above statement it is sufficient to show that
\be\label{eq:SigmaK} \delta\tilde \mu=\int_{M_1} {\Sigma }_i\wedge
\delta K^i-\int_{M_2} {\Sigma }_i\wedge \delta K^i \ee for some
functional $\tilde \mu$.

The isolated horizon boundary condition imply that the only
variations of the fields allowed on the horizon $\Delta$ are
infinitesimal $SL(2,\mathbb{C})$ gauge transformations and
diffeomorphisms, namely: \ba\label{fasy}
&&\delta{e}=\delta_{\alpha}{e}+\delta_v{e} \n\\
&&\delta {A^+}=\delta_{\alpha}{A^+}+\delta_v{A^+}\n\\
&&\delta{K}=\delta_{\alpha}{K}+\delta_v{K} \ea where $\alpha:
M\rightarrow sl(2,\mathbb{C})$ and $v$ is a vector field tangent to
$H$. Under such transformations we have: \ba\label{eq:Gauss
Variations} && \delta_{\alpha}e^i=[\alpha,
e]^i~~~~~~~~~\delta_{\alpha}\Sigma^{+i}=[\alpha, \Sigma^+]^i
~~~~~~~~~\delta_{\alpha}A^{+i}=-d_{A^+} \alpha^i
\ea \vspace{-24pt}
\ba
&& \delta_v e^i=v\inter de^i + d(v\inter e^i)
\n\\
&& \delta_v A^{+i} =v\inter F^i(A^+) + d_{A^+}(v\inter A^{+i})
\n\\
&& \delta_v \Sigma^{^+i} =d_{A^+}(v\inter \Sigma^+)^i-[v\inter
A^+,\Sigma^+]^i\label{eq:Diffeo Variations} \ea where $(v\inter
\omega)_{b_1\cdots b_{p-1}}\equiv v^a\omega_{ab_1\cdots b_{p-1}}$
for any $p$-form $\omega_{b_1\cdots b_p}$, and in the last line we
used the Gauss law.
Let us also recall the some useful relations: \be\label{eq:AB=-AB}
A\wedge v\inter B=- v\inter A\wedge B \ee for any 2-form $A$ and
1-form $B$ on a 2-manifold or any 2-form $A$ and 2-form $B$ on a
3-manifold, while \be\label{eq:AB=AB} A\wedge v\inter B= v\inter
A\wedge B \ee for any 1-form $A$ and 2-form $B$ on a 2-manifold or
any 1-form $A$ and 3-form $B$ on a 3-manifold.
%
We used the notation
$\Gamma^i=-\frac{1}{2}\epsilon^i\,_{jk}\omega^{jk}$ when working
with the torsion free connection compatible with the triad $e^i$. In
this notation the torsion free condition (Cartan's first structure
equation) takes the form \be\label{eq:Cartan}
de^i=-\epsilon^i\,_{jk}\Gamma^j\wedge e^k \ee which  implies \be
F^i(\Gamma)=d\Gamma^i+\frac{1}{2}\epsilon^i\,_{jk}\Gamma^j\wedge\Gamma^k\,
\ee A very important property of the spin connection \cite{lqg1}
compatible with $e^i$ is that \be\label{idee}
 \int_{M} \Sigma_i\wedge
\delta \Gamma^i = \int_{H} - e_i \wedge \delta e^i \ee This identity
allows one to rewrite (\ref{eq:Potential}) as
\ba\label{ss}\delta\mu-\i \sD(\delta)=\int_{M_1} {\Sigma }_i\wedge
\delta K^i-\int_{M_2} {\Sigma }_i\wedge \delta K^i \ea where \be
\sD(\delta)=\int_{H_1-H_2} e_i\wedge \delta e^i- \int_\Delta
\Sigma^+_i\wedge \delta A^{+i} \ee More specifically, we can
evaluate the previous equation on the allowed variations on the
boundary. For gauge transformations we get \ba \i
\sD(\delta_\alpha)&=&\i\int_{H_1-H_2} e_i \wedge[\alpha, e]^i
+\i \int_\Delta d(\Sigma^+_i \alpha^i)-d{\Sigma^+}_i \alpha^i+{\Sigma^+}_i\wedge \epsilon^i\,_{jk}{A^+}^j\alpha^k \n\\
&=&\i\int_{H_1-H_2} e_i \wedge \epsilon^i\,_{jk}\alpha^je^k
+\i \int_\Delta d({\Sigma^+}_i\alpha^i)-d{\Sigma^+}_i \alpha^i-\epsilon_{ijk}{A^+}^j\wedge{\Sigma^+}^k \alpha^i \n\\
&=&-\i\int_{H_1-H_2} {\Sigma }_i\alpha^i +\i
\int_{H_1-H_2}{\Sigma}_i\alpha^i -\i\int_\Delta d_{A^+}{\Sigma^+
}_i\, \alpha^i=0 \ea where in the last line we have used the Gauss
law. Therefore, for gauge transformations on the horizon, we have
proven the result (\ref{eq:SigmaK}). Let us now analyze the
diffeomorphisms. Analogously to the gauge transformations case, we
have \ba \i \sD(\delta_v) &=&\i\int_{H_1-H_2} e_i \wedge
\left(v\inter de^i + d(v\inter e^i)\right)
-\i \int_\Delta {\Sigma^+}_i\wedge \left(v\inter F^i(A^+) + d_{A^+}(v\inter A^{+i})\right) \n\\
&=&\i\int_{H_1-H_2}2v\inter e_i de^i-d(e_i v\inter e^i)
-\i\int_\Delta d({\Sigma^+}_iv\inter A^{+i})-d_{A^+} {\Sigma^+}_i\,v\inter A^{+i} \n\\
&=&\i\int_{H_1-H_2}{\Sigma}_iv\inter\Gamma^i-\i\int_{H_1-H_2}{\Sigma}_iv\inter A^{+i} \n\\
&=&\int_{H_1-H_2}{\Sigma}_iv\inter
K^i=-\int_{H_1-H_2}v\inter{\Sigma}_i\wedge K^i \ea where in the
first line we have used the vector constraint, in the second one the
Gauss law and in the third one the Cartan equation. Therefore, in
the case of diffeomorphisms for the variations on the horizon, the
result (\ref{eq:SigmaK}) holds as long as \be\label{SigmaK=0}
v\inter{\Sigma }_i\wedge K^i=0 \ee If the previous relation is
satisfied, {\em i.e.}, if the isolated horizon is static. From
(\ref{eq:SigmaK}) it follows that the conserved symplectic form for
gravity can be written as: \be\label{eq:Symplectic Form1}
\kappa\,\Omega(\delta_1,\delta_2)=\int_M \delta_{[1}\Sigma^i\wedge
\delta_{2]} K_i \ee where $M$ is a Chauchy surface representing
space and $\delta_1,\delta_2\in T_p\Gamma$, {\em i.e.}, they are
vectors in the tangent space to the phase-space $\Gamma$ at the
point $p$. The symplectic form above is manifestly real and has no
boundary contribution.

\subsection{Gauge Symmetries}\label{GSS}

The gauge symmetry content of isolated horizon system is
characterized by the degenerate directions of the presymplectic
structure. In fact we now show how tangent vectors $\delta_\alpha,
\delta_{v}\in T_p\Gamma$ of the form
\begin{eqnarray} &&\nonumber
\delta_{\alpha} \Sigma=[\alpha,\Sigma],\ \
\delta_{\alpha}K=[\alpha,K]; \\ && \delta_{v} \Sigma= \Lie_v
\Sigma=v\inter d\Sigma+d(v\inter\Sigma), \ \ \delta_{v}K= \Lie_v
K=v\inter dK+d(v\inter K) \label{ttta}\end{eqnarray} for $\alpha: M
\rightarrow \frak{su}(2)$ and $v \in \Vect(M)$ tangent to the
horizon, are degenerate directions of $\Omega_{M}$ if an only if the
isolated horizon is static.

The proof given here follows exactly the one presented in
\cite{Spherically}. However, our analysis here applies to generic
static isolated horizons instead of simply Type I isolated horizons.
We start with the $SU(2)$ transformations $\delta_\alpha$, and we
get \ba \label{su2} && \kappa
\Omega_{M}(\delta_\alpha,\delta)=\int_{M} [\alpha,\Sigma]_i\wedge
\delta K^{i}- \delta \Sigma_i\wedge [\alpha,K]^i= \int_{M}
\delta({\epsilon_{ijk}\alpha^j \Sigma^k\wedge K^i)})=0 \ea where we
used the Gauss constraint $\epsilon_{ijk}\Sigma^k\wedge K^i=0$. In
order to treat the case of the infinitesimal diffeomorphisms tangent
to the horizon $H$ it will be convenient to first write the form of
the vector constraint $V_a$ in terms of $\Sigma-K$ variables
\cite{tate}. We have \ba v\inter V=dK^i\wedge
v\inter\Sigma_i+v\inter K^i\ d\Sigma_i\approx0 \ea variations of the
previous equation yields \ba v\inter \delta V&=&d(\delta K)^i \wedge
v\inter\Sigma_i+dK^i\wedge v\inter\delta\Sigma_i+v\inter \delta K^i\
d\Sigma_i+v\inter K^i\ d(\delta\Sigma)_i \nonumber \\ &=&
v\inter\Sigma_i\wedge d(\delta K)^i-\delta\Sigma_i\wedge v\inter
dK^i+  v\inter d\Sigma_i \wedge \delta K^i+ d(\delta\Sigma)_iv\inter
K^i =0 \ea where in the second line we have put all the $K$'s to the
right, and modified the second and third terms using the identities
(\ref{eq:AB=-AB})--(\ref{eq:AB=AB}).
We are now ready to show that $\delta_v$ is a null direction of
$\Omega_M$. Explicitly: \ba &&\nonumber \kappa
\Omega_{M}(\delta_v,\delta)=\int_{M} (v\inter
d\Sigma+d(v\inter\Sigma))_i\wedge \delta K^{i}- \delta
\Sigma_i\wedge (v\inter dK+d(v\inter K))^i \\
&&\nonumber =  \int_{M} v\inter d\Sigma_i\wedge \delta
K^{i}+d(v\inter\Sigma)_i\wedge \delta K^{i}- \delta \Sigma^i\wedge
v\inter dK_i- \delta
\Sigma_i \wedge d(v\inter K)^i \\
&&\nonumber =  \int_{M}\underbrace{ v\inter d\Sigma_i\wedge \delta
K^{i}+v\inter\Sigma_i\wedge d(\delta K)^{i}- \delta \Sigma^i\wedge
v\inter dK_i+d(\delta \Sigma)^i \wedge v\inter K_i}_{v\inter \delta
V=0} \\ && +\int_{\partial M}v\inter\Sigma_i\wedge \delta K^{i}
-\delta \Sigma_i \wedge v\inter K^i=\int_{\partial
M}\delta(v\inter\Sigma_i\wedge K^{i})=0  \label{diffh}\ea where in
the last line we have used again the identity (\ref{eq:AB=-AB}), the
fact that $v$ is tangent to $H$, and the IH staticity condition
(\ref{stati})---from where (\ref{k1}) follows---implying that
$v\inter \Sigma_i\wedge K^i=0$ when pulled back to the horizon.
$\square$ \vskip.2cm

\subsection{The Conserved Symplectic Structure for Non-Static Isolated Horizons}\label{ns}

We have seen that the symplectic structure (\ref{eq:Symplectic Form1}) is conserved if and only if the isolated horizon is static.
The last equation of the previous subsection gives us a hint of what is the source of the problem. The source of all the difficulties is that, in the non static case, (\ref{eq:Symplectic Form1}) 
breaks diffeomorphisms invariance. 
This is indeed an important point as it implies that the usual techniques for quantization and calculation of entropy are not applicable for 
non static isolated horizons which contain in particular the Kerr black hole.
 There are two ways out of this difficulty:
\begin{enumerate}
\item To declare that diffeomorphism are not gauge symmetries in the non-static case and modify the definition of the phase space
allowing variations of fields at the horizon which are only pure $SU(2)$ gauge transformations (i.e. what would correspond to  $v=0$ in (\ref{fasy})).
\item To restore diffeomorphism invariance at the horizon by the inclusion of new field degrees of freedom.
\end{enumerate}

The first possibility appears as the simplest way out, and it is indeed a natural one often used in other contexts. For instance this is done in the construction of the phase space of asymptotically flat spacetimes where the boundary conditions reduce the gauge symmetries to global symmetries allowing in this way the definition of non trivial charges such us linear and angular momentum at infinity.  However, this approach is not viable in the present context where one intends to define a model capable of accounting for the degrees of freedom leading to black hole entropy. The reason is that diffeomorphism invariance is a central ingredient for the consistency of the quantum theory describing the degrees of freedom at the horizon and their relationship with the bulk loop quantum gravity degrees of freedom. The lack of diffeomorphism invariance in the phase space of the isolated horizon makes the usual program inapplicable.

Consequently, here we make a small step in the direction of the second possibility and propose a concrete extension of the phase space restring the conservation of the
symplectic structure with a simple extension of (\ref{fasy}).  
As it is well known and used in other contexts \footnote{There is a nice analogy in a simpler context: when one couples particles to $2+1$ gravity one initially breaks diffeomorphism and gauge invariance at points. The degrees of freedom of the coupled particles are naturally defined by the requirement that the gauge symmetries are restored which necessitates the extension of the field content at the position of the particles.}, 
diffeomorphism invariance can be restored by introducing additional degrees of freedom on the horizon. Concretely, one should replace (\ref{eq:Symplectic Form1}) by
\be\label{nonsta}
\kappa\,\Omega_{ns}(\delta_1,\delta_2)=\int_M \delta_{[1}\Sigma^i\wedge \delta_{2]} K_i+\int_{H} \delta_{[1}J\wedge \delta_{2]}\phi
 \,,
\ee
where $J$ is a $2$-form field and $\phi
$ a scalar field which under infinitesimal diffeomorphisms tangent to $H$ transform 
in the usual way, namely
\be
\delta_v J=d(v\inter J)\ \ \ \delta_v\phi
=v\inter d\phi
.
\ee
We see immediately that 
\ba
\kappa\,\Omega_{ns}(\delta_v,\delta)&=&\int_{\partial M}\delta(v\inter\Sigma_i\wedge K^{i})+d(v\inter J)\wedge \delta \phi
 -\delta J\wedge v\inter d\phi
\n\\ &=&\int_{\partial M}\delta(v\inter\Sigma_i\wedge K^{i}- J\wedge v\inter d \phi
 ),
\ea
which would vanish as desired if we imposed the restriction
\be\label{angy}
v\inter\Sigma_i\wedge K^{i}- J\wedge v\inter d \phi
=0
\ee
on the new fields.  The previous constraint is nothing else but the diffeomorphism constraint of the 
new extended system. In the axisymmetric case (e.g. the Kerr solution) $\phi
=\varphi$ and the previous equation implies
$J=\Sigma_i\wedge K_{\varphi}^{i}$ which is precisely the angular momentum density, indeed for Kerr \cite{Chandra}, one has
\be
M a=\int_{H}J=\int_{H} \Sigma_i\wedge K_{\varphi}^{i},\ee
where $M$ is the mass and $a$ is the angular momentum per unit mass of the Kerr spacetime. 

This trick allows us for restoring diffeomorphism invariance but the question remains: is the new symplectic structure (\ref{nonsta}) conserved for an non static isolated horizon?
The answer is in the affirmative as the following calculation shows. We can simply adapt the conservation proof given in section \ref{Conservation} to this case. Indeed, if we start from equation (\ref{ss}) which is valid for generic isolated horizons
and we add on both sides the same term 
\[\int_{H_1-H_2} J\wedge \delta \phi
\] we get \ba
\delta\mu-\i
\sD_{ns}(\delta)=\int_{M_1} {\Sigma
}_i\wedge \delta K^i + \int_{H_1} J\wedge \delta \phi
 -\int_{M_2} {\Sigma
}_i\wedge \delta K^i -\int_{H_2} J\wedge \delta \phi
,
\ea
where
\be
\sD_{ns}(\delta)=\int_{H_1-H_2} (e_i\wedge \delta e^i+J\wedge \delta\phi
)- \int_\Delta {\Sigma^+}_i\wedge \delta A^{+i}.
\ee
As $\delta_{\alpha} J=0=\delta_{\alpha} \phi
$, for gauge transformations we get  
\ba
\i
\sD_{ns}(\delta_\alpha)=0 .\ea
Let us now analyze the diffeomorphisms;
\ba
\i
\sD_{ns}(\delta_v) =\int_{H_1-H_2}-v\inter{\Sigma
}_i\wedge K^i+J\wedge v\inter d\phi=0
,
\ea
according to (\ref{angy}). Therefore,  the new symplectic structure 
(\ref{nonsta}) is conserved and diffeomorphism invariance at the boundary is restored with the addition of the new fields $(J,\phi)$.
This extended system is now suitable to be quantized in consistency with the bulk theory. The understanding of the BH entropy calculation for rotating black holes
could be based on this formulation.
The quantization and interpretation of this new system is clearly a more difficult problem that we hope to study in more detail in the future.

\subsection{The Conserved Symplectic Structure in Terms of Real Connection Variables}
We now want to introduce the Ashtekar-Barbero variables \be
A^i_a=\Gamma^i_a+\beta K^i_a\, \ee where
$\Gamma^i=-\frac{1}{2}\epsilon^{ijk}\omega_{jk}$ and $\beta$ is the
Immirzi parameter. We can write the symplectic potential
corresponding to (\ref{eq:Symplectic Form1}) as \ba\n \kappa\,
\Theta(\delta)&=&\frac{1}{\beta} \int_{M} \Sigma_i\wedge
\delta(\Gamma^i+\beta K^i)-\frac{1}{\beta} \int_{M} \Sigma_i\wedge
\delta \Gamma^i\\ &=& \frac{1}{\beta} \int_{M} \Sigma_i\wedge
\delta(\Gamma^i+\beta K^i)+\frac{1}{\beta} \int_{H} e_i\wedge \delta
e^i \ea where in the last line we have used (\ref{idee}). In terms
of the Ashtekar-Barbero connection the symplectic structure
(\ref{eq:Symplectic Form1}) takes the form: \ba\label{eq:Symplectic
Form2} \kappa\,\Omega_M(\delta_1,\delta_2) &=&\frac{1}{\beta}\int_M
\delta_{[1}\Sigma^i\wedge \delta_{2]} A_i- \frac{1}{\beta}\int_{H}
\delta_{[1}e^i\wedge \delta_{2]} e_i\, \ea where $H$ is the boundary
of $M$.

Let us now comment on the nature of this result. We have shown that
in the presence of a static isolated horizon the conserved
pre-symplectic structure is the usual one when written in terms of
vector-like (or Palatini) variables. When we write the
pre-symplectic structure in terms of Ashtekar-Barbero connection
variables in the bulk, the pre-symplectic structure acquires a
boundary term at the horizon of the simple form \cite{Spherically,
wi} \be\label{bb} \kappa\,\Omega_H(\delta_1,\delta_2)=
\frac{1}{\beta}\int_{H} \delta_{[1}e^i\wedge \delta_{2]} e_i\, \ee A
first observation is that, as shown in \cite{Spherically}, this
implies the kind of non-commutativity of flux variables that is
compatible with the use of the holonomy-flux algebra as the starting
point for quantization. This (completely continuum classical
analysis) reinforces in this sense the importance of the kinematical
quantization at the basis of the definition of loop quantum gravity
\cite{lost1,lost2}. A second observation is that the degrees of
freedom at the horizon are encoded in the pull back of the triad
fields $e^i$ on the horizon subjected to the obvious constraint
\be\Sigma^i_H=\Sigma^i_{Bulk} \ee which are three first class
constraints---as it follows from (\ref{bb})---for the six
unconstrained phase space variables $e^i$. Therefore, one would
expect that (as in the Type I case \cite{prl,Spherically}) all of
the boundary degrees of freedom are fixed by the bulk degrees of
freedom. However, this turn out not to be that case if one works
with triad fields as fundamental. The difficulty appears in the
quantum theory where degenerate geometries are admitted. This can be
visualized by concentrating on the special case
$\Sigma^i_H=0$---that would correspond in the quantum theory to
points where there is no bulk spin-network puncture. It is easy to
see that the constraints $\Sigma^i_H=0$  does not kill all the local
degrees of freedom in $e^i$ as there is a non-trivial moduli space
of degenerate $e^i$'s that would naively lead to an infinite
entropy \footnote{This can be made precise
by studying in more detail the quantization of the $e^i$ fields on $H$. This more detailed analysis offers interesting possibilities beyond the scope of this work. 
Results are going to be reported elsewhere \cite{laurent}.}.

This difficulty is clearly related to the choice of continuum
variables used for the parametrization of the boundary phase space.
For instance, in the spherically symmetric case, the degrees of
freedom are encoded instead in a connection $A^i$ and the analog of
the constraints $\Sigma^i_H=0$ (where there are no bulk punctures)
are $F^i(A)=0$. The dimensionality of both the unconstrained phase
space and constraint surfaces are the same as in the treatment based
on triads; however, the constraint $F^i(A)=0$ completely annihilates
the local degrees of freedom at places where there are no
punctures---as it implies that $A=gdg^{-1}$ (pure gauge)
locally---rendering the entropy finite.  This motivates the use of
connection variables to describe (\ref{bb}). One of the results of
this paper is to show that this can indeed be achieved for generic
IH by the introduction of a pair of connection variables \be
\label{coco} A^i_\gamma=\Gamma^i+\gamma e^i\ \ \  {\rm and}\ \ \
A^i_\sigma=\Gamma^i+\sigma e^i\ee  in terms of which the boundary
term of the conserved symplectic form (\ref{bb}) becomes
\ba\label{boundy} \kappa\beta\,\Omega_{\va
H}(\delta_1,\delta_2)=\frac{1}{\sigma^2-\gamma^2}\int_H\delta_{[1}A_\gamma^i\wedge
\delta_{2]} A_{\gamma i}-
\frac{1}{\sigma^2-\gamma^2}\int_H\delta_{[1}A_{\sigma}^i\wedge
\delta_{2]} A_{\sigma i}\, \ea The proof of this statement is
presented in the following subsection. From the IH boundary
conditions, through the relations (\ref{UNO}) and
(\ref{eq:KK=lambda_0 Sigma}), Cartan equations, and the definitions
(\ref{coco}), the following relations for the new variables follow
\ba\label{newy}
F^i(A_\gamma)
&=&\Psi_2\Sigma^i+\frac{1}{2}(\gamma^2+c)\Sigma^i\n\\
F^i(A_\sigma)
&=&\Psi_2\Sigma^i+\frac{1}{2}(\sigma^2+c)\Sigma^i \ea This means
that there is a two-parameter family of equivalent classical
descriptions of the system that in terms of triad variables is
described by (\ref{bb}) (we will see in the sequel that the two
parameter freedom reduces indeed to a single one when additional
consistency requirements are taken into account). The appearance of
these new parameters $\sigma$ and $\gamma$ is strictly related with
the introduction of the $SU(2)$ connection variables (as was already
observed in \cite{Spherically}). In this sense the situation is
completely analogous to the one leading to the appearance of the
Immirzi parameter when going from vector (Palatini) variables to
Ashtekar-Barbero variables in the parametrization of the phase space
of general relativity in the bulk.

In the quantum theory, at points where there are no punctures from
the bulk, the two connections are subjected to the six first class
constraints $F^i(A_\gamma)=0=F^i(A_{\sigma})$ implying the absence
of local degrees of freedom at these places. The new variables
resolve in this way the difficulty that we encountered in the
treatment in terms of the triads $e^i$. In addition, the connection
fields $A_{\gamma}$ and $A_{\sigma}$  are described by Chern-Simons
symplectic structures respectively, which will allow the use of some
of the standard techniques, applicable to Type I isolated horizons,
for the quantization of arbitrary static isolated horizons.

\vspace{12pt} \noindent {\bf Remark:} Using the well known
relationship between Chern-Simons theory and 2+1 gravity
\cite{CSgravity1,CSgravity2} it is possible to rewrite
(\ref{boundy}) in terms of 2+1 gravity like variables: an $SU(2)$
connection and a `triad' field. However, the coupling constraints
(\ref{newy}) become more cumbersome in the prospect of quantization.

\subsection{Equivalence between the Triad and Connection Parametrizations of the Boundary Symplectic Structure}\label{app}

In this section  we present the proof of the validity of Equation
(\ref{boundy}). The key point---from which (\ref{boundy}) follows
directly---is to show that phase space one-form $\Theta_0(\delta)$
defined by \be \Theta_0(\delta) \equiv  \int_{H}  e_i \wedge \delta
e^i - \frac{1}{\sigma^2-\gamma^2}\int_{H}\left( A_{\gamma i}\wedge
\delta A^i_{ \gamma}-A_{\sigma i}\wedge \delta A^i_{ \sigma}\right)
\ee is indeed closed.

\vspace{12pt} {\noindent \em Proof:}
 Let us denote by
\[{\frak d}\Theta_0 (\delta_1,\delta_2)=\delta_1(\Theta_0(\delta_{2}))-\delta_2 (\Theta_0(\delta_{1}))\]
the exterior derivative of $\Theta_0$.  For infinitesimal $SU(2)$
transformations we have \be \delta_{\alpha}e=[\alpha,e]\ \ \ \ \ \
\delta_{\alpha}A_{\gamma(\sigma)}=-d_{A_{\gamma(\sigma)}} \alpha
\label{su2cc} \ee from which it follows: \ba
 \nonumber {\frak d}\Theta_0(\delta, \delta_{\alpha})&=&
\int_{H}\delta_{}e^i\wedge \delta_{\alpha}e_i
-\frac{2}{\sigma^2-\gamma^2}\int_{H}\left(\delta_{}A_{\gamma i}\wedge \delta_{\alpha} A^i_{ \gamma}-\delta_{}A_{\sigma i}\wedge \delta_{\alpha} A^i_{ \sigma}\right) \n\\
&=&\int_{H}2\delta e^i\wedge \epsilon_{ijk}\alpha^je^k
+\frac{2}{\sigma^2-\gamma^2}\int_{H}\left( \delta A_{\gamma i}\wedge
d_{A_\gamma} \alpha^i-
\delta A_{\sigma i}\wedge d_{A_\sigma} \alpha^i\right) \n\\
&=&\int_{H}2\delta e^i\wedge e^k\alpha^j\epsilon_{ijk}
+\frac{2}{\sigma^2-\gamma^2}\int_{H} \left( (d_{A_\gamma}\delta
A_{\gamma i} )  \alpha^i-
(d_{A_\sigma}\delta A_{\sigma i})  \alpha^i\right) \n\\
&=&\int_{H}-2\epsilon_{ikj}\delta e^i\wedge e^k\alpha^j
+\frac{2}{\sigma^2-\gamma^2}\int_{H}\left( \delta F_i(A_\gamma  )
\alpha^i-
\delta F_i(A_\sigma  )   \alpha^i\right) \n\\
&=&\int_{H} \delta(-\Sigma^i+\frac{2}{\sigma^2-\gamma^2}
(F_i(A_\gamma  ) -F_i(A_\sigma  ) ))\alpha_i\, \ea where in the
third line we have integrated by parts. Therefore, from
(\ref{newy}), we have that \be {\frak d}\Theta_0(\delta,
\delta_{\alpha})=0 \ee For tangent diffeomorphisms on $H$, using
Equation (\ref{eq:Diffeo Variations}), we have \ba \nonumber {\frak
d}\Theta_0(\delta, \delta_{v})&=&\int_{H} 2\delta e^i\wedge
\delta_ve_i -\frac{2}{\sigma^2-\gamma^2
} \int_{H}\left(\delta A_{\gamma i}\wedge \delta_v A^i_{ \gamma}-\delta A_{\sigma i}\wedge \delta_v A^i_{ \sigma}\right) \n\\
&=&\int_{H}2v\inter \delta e_i\wedge de^i + 2\delta d e_i \wedge v\inter e^i  \n\\
&-&\frac{2}{\sigma^2-\gamma^2
}\int_{H}\left(\delta \Gamma_i \wedge v\inter (F^i(A_{\gamma})-F^i(A_{\sigma}))+(\gamma-\sigma)\delta e_i\wedge v\inter (F^i(A_{\gamma})-F^i(A_{\sigma}))\right)\n\\
&-&\frac{2}{\sigma^2-\gamma^2
}\int_{H}(\delta F_i(A_{\gamma})v \inter A^i_\gamma -\delta F_i(A_{\sigma})v \inter A^i_\sigma) \n\\
&=&\int_{H}2\delta ( de^i \wedge v\inter e_i)
-\frac{2}{\sigma^2-\gamma^2
}(\gamma-\sigma)\int_{H}\underbrace{\delta \left(e_i\wedge v\inter (F^i(A_{\gamma})-F^i(A_{\sigma}))\right)}_{\propto\delta(e_i\wedge v\inter \Sigma^i ) =0}\n\\
&-&\int_{H}\left(\delta \Gamma_i \wedge v\inter \Sigma^i\right)
-\int_{H}\left(\delta \Sigma_i \wedge v\inter \Gamma^i\right) \n\\
&=&\int_{H}-2\delta ( \epsilon^i\,_{jk}\Gamma^j\wedge e^k \wedge
v\inter e_i)
-\int_{H}\delta ( \Gamma^i\wedge v\inter \Sigma_i ) \n\\
&=&\int_{H}\delta ( \Gamma^i\wedge v\inter \Sigma_i )
-\int_{H}\delta ( \Gamma^i\wedge v\inter \Sigma_i )=0\, \ea where in
addition to integrating by parts and using that $\partial H=0$, we
have used Cartan's structure equation (\ref{eq:Cartan}), the fact
that $e^1$ is orthogonal to $H$ (that is why the expression below
the underbracket is zero), and the identities $ A\wedge v\inter B=-
v\inter A\wedge B$ valid for any 2-form $A$ and 1-form $B$ on a
2-manifold or any 2-form $A$ and 2-form $B$ on a 3-manifold.
$\square$

The previous statement implies that the boundary symplectic form can
be rewritten as \be \beta\kappa\,\Omega(\delta_1,\delta_2)=\int_M
\delta_{[1}\Sigma^i\wedge \delta_{2]} A_i-
\frac{1}{(\sigma^2-\gamma^2)}\int_{H}\left( \delta_{[1}A_{\gamma
i}\wedge \delta_{2]} A^i_{ \gamma}-\delta_{[1}A_{\sigma i}\wedge
\delta_{2]}  A^i_{ \sigma}\right) \ee From the previous equation we
conclude that the boundary term is given by the simplectic structure
of two Chern-Simons theories  with levels \be\label{eq:Level}
k_\gamma=-k_\sigma=\frac{8\pi}{ (\sigma^2-\gamma^2)\kappa\beta} \ee
%

\section{Quantization}\label{Quantization}

{ The form of the symplectic structure motivates one to handle the
quantization of the bulk and horizon degrees of freedom separately.
We first discuss the bulk quantization. As in standard LQG
\cite{lqg1,lqg2,lqg3,lqg4} one first considers (bulk) Hilbert spaces
$\sH^B_\gamma$ defined on a graph $\gamma \subset M$ and then takes
the projective limit containing the Hilbert spaces for arbitrary
graphs. Along these lines let us first consider $\sH^{\va
B}_{\gamma}$ for a fixed graph $\gamma \subset M$ with end points on
$H$, denoted $\gamma\cap H$. The quantum operator associated with
$\Sigma$ is the flux operator of LQG that can be written as
%
%
\begin{equation}
\label{gammasigma} \epsilon^{cd}\Sigma^i_{cd}(x) = 16 \pi G\hbar
\beta \sum_{p \in \gamma\cap H} \delta(x,x_p) J_b^i(p)
\end{equation}
where $J^i_{b}(p)$ are $SU(2)$ infinitesimal generators, such that
$[J_b^i(p),J_b^j(p)]=\epsilon^{ij}_{\ \ k} J_b^k(p)$, acting at each
$p\in\gamma\cap H$ and $\gamma \subset M$ is a graph with end points
$p$ on $H$ (for more details see \cite{lqg1,lqg2,lqg3,lqg4}). The
index $b$ in $J_b(p)$ {is not a space-time index, but simply} stands
for bulk. Consider a basis of $\sH^{{\va B}}_{\gamma}$ of
eigenstates of both $J(p)\cdot J(p)$ as well as $J^3(p)$ for all
$p\in \gamma\cap H$ with eigenvalues $\hbar^2 j_p(j_p+1)$ and $\hbar
m_p$ respectively. These states are spin network states, here
denoted $|\{j_p,m_p\}_{\va 1}^{\va n}; {\van \cdots} \rangle$, where
$j_p$ and $m_p$ are the spins and magnetic numbers labeling the $n$
edges puncturing the horizon at points $x_p$ (other labels are left
implicit). They are also eigenstates of the horizon area operator
$a_{\va H}$ \be\label{area} a_{\va H}|\{j_p,m_p\}_{\va 1}^{\va n};
{\van \cdots} \rangle=8\pi\beta \ell_p^2 \,
\sum_{p=1}^{n}\sqrt{j_p(j_p+1)} |\{j_p,m_p\}_{\va 1}^{\va n}; {\van
\cdots} \rangle \ee}

Now, following Witten's prescription to quantize the two
Chern-Simons theories with punctures \cite{Witten}, we introduce:
\be\label{eq:Classical} \frac{{k_{\gamma}}}{ 4\pi\hbar
}F^i(A_{\gamma})=J_{\gamma}^i(p)\ \ \ \ \frac{{k_{\sigma}}}{
4\pi\hbar }F^i(A_{\sigma})=J_{\sigma}^i(p)\ \ee where \be
\label{level}
{k}_{\gamma}=-k_{\sigma}=\frac{8\pi}{(\sigma^2-\gamma^2)\kappa\beta}\ee
are the two levels of the two $SU(2)$ Chern-Simons theories involved
in our model. If we do so we can now write the constraints as
(recall (\ref{newy})): \ba\label{eq:J+J-}
J^i_{\gamma}(p)&=&\frac{2\Psi_2 +(\gamma^2+c)}{(\sigma^2-\gamma^2)}\ J^i_b(p) \n\\
J^i_{\sigma }(p)&=&
-\frac{2\Psi_2+(\sigma^2+c)}{(\sigma^2-\gamma^2)} \ J^i_b(p) \ea So
now from (\ref{eq:J+J-})  we obtain the constraint \ba\label{gaussy}
&&D^i(p)=J^i_b(p)+J_\gamma^i(p)+J_\sigma^i(p)=0 \ea  plus the
constraint \be
\label{disty}C^i(p)=J_{\gamma}^i(p)-J_\sigma^i(p)+\frac{ 2d+
(\sigma^2+\gamma^2)}{(\sigma^2-\gamma^2)} J^i_b(p)=0 \ee The
constraints $C^i(p)=0$ will provide the definition of $d=2\Psi_2+c$
in the quantum theory (see Equation~(\ref{gauss})). This point will
be clarified at the end of the following subsection. From now on we
shall set $\hbar=1$ for notational convenience.

We can quantize the boundary theory following Witten's prescription.
In fact, the Hilbert space of the boundary model is that of two
Chern-Simons theories associated with a pair of spins
$(j^\gamma_p,j^\sigma_p)$ at each puncture. More precisely,
\be\label{eq:Hilbert Space} \mathscr{H}^{CS}_k (j^\gamma_1\cdots
j^\gamma_n) \otimes\mathscr{H}^{CS}_k(j^\sigma_1\cdots
j^\sigma_n)\subset {\rm Inv}(j^\gamma_1\otimes\cdots \otimes
j^\gamma_n) \otimes {\rm Inv}(j^\sigma_1 \otimes\cdots \otimes
j^\sigma_n)\, \ee {The Hilbert space of CS theory with given
punctures on the sphere can be thought of as the intertwiner space
of the quantum deformation of $SU(2)$ denoted $U_q(su(2))$. The
inclusion symbol in the previous expression means that the later
space is isomorphic to a subspace of classical $SU(2)$ intertwiner
space. This is due to the fact that, in this isomorphism, the spins
associated to the $CS$ punctures cannot take all values allowed by
the representation theory of $SU(2)$, but are restricted by the
cut-off $k/2$ related to the deformation parameter by
$q=\exp({\frac{\i \pi}{k+2}})$ where $k$ is the Chern-Simons level.}
%
%

The operators associated to $J^i_\gamma(p)$ and $J^i_\sigma(p)$, on
the other hand, describe the spins of the pair of CS defects at the
punctures. They are observables of the boundary system with which
the spins $j^{\gamma}_p$ and $j^\sigma_p$ are associated. The theory
is topological  which means in our case that non-trivial d.o.f. are
only present at punctures. With all this we can now impose
(\ref{gaussy}) which, at a single puncture, requires invariance
under $SU(2)$ local transformations $$\delta J_{A}^j =[\alpha_i
D^i,J^j_{A}]=\epsilon^{ij}_{\ k}\alpha_iJ_{A}^k $$ where the
subscript  $A$ here stands for $\gamma$, $\sigma,$ and $b$
respectively. Equivalently, the constraint  (\ref{gaussy})  requires
the quantum state to be proportional to the singlet state with zero
total $SU(2)$ charge: zero total angular momentum. More precisely,
the constraint $D^i(p)=0$ simply requires that \be {\rm
Inv}(j_p\otimes j_p^\gamma\otimes j_p^\sigma)\not=\emptyset
\label{tritri} \ee at each puncture $p$.


\subsection*{Learning from the Restriction to Type I Isolated Horizons}\label{2-1}

Let us now show how, within this new approach, we are able to
recover the same picture of the spherically symmetric case as
studied in \cite{Spherically}. We will show how this will allow us
to reduce from two to one the number of free parameters describing
the system. For a spherically symmetric IH we can replace in
(\ref{eq:J+J-}) $\Psi_2$ and $c$ with their constant classical
values: $\Psi_2=-\frac{2\pi}{a_{\va H}}$ and $c=\frac{2\pi}{a_{\va
H}}$ (see \cite{Spherically}). If we do so, we obtain the two sets
of constraints: \ba &&D^i(p)\equiv
J_b^i(p)+J_\gamma^i(p)+J_\sigma^i(p)=0~~~~~~~~~~~~~C^i(p)\equiv
J_\gamma^i(p)-J_\sigma^i(p)+\alpha J_b^i(p)=0  \label{eq:J+,J-
Spherical} \, \ea where \[\alpha\equiv\frac{
(\sigma^2+\gamma^2)-\frac{4\pi}{a_{\va H}}}{\sigma^2-\gamma^2} \] {
The form of the symplectic structure as in (\ref{bb}) and
(\ref{boundy})---together with Equations
(\ref{eq:Classical})---implies the commutation relations
$[J_{\gamma}^i(p),J_{\gamma}^j(p)]=\epsilon^{ij}_{\ \ k}
J^k_{\gamma}(p)$,
$[J_{\sigma}^i(p),J_{\sigma}^j(p)]=\epsilon^{ij}_{\ \ k}
J^k_{\sigma}(p)$, and $[J_{b}^i(p),J_{b}^j(p)]=\epsilon^{ij}_{\ \ k}
J^k_{b}(p)$ from which the constraint algebra that follows is:}
\ba &&[C^i(p),C^j(p^{\prime})]=\epsilon^{ij}\!_k\ (J_\gamma
^k(p)+J_\sigma
^k(p)+\alpha^2J_b^k(p) )\, \delta_{pp'}\label{eq:C-C}\\
&&[C^i(p), D^j(p')]=\epsilon^{ij}\!_k \ C^k(p)\, \delta_{pp'} \label{eq:C-D}\\
&&[D^i(p),D^j(p')]= \epsilon^{ij}\!_k \ D^k(p) \,
\delta_{pp'}\label{eq:D-D} \ea We can thus see that, by setting
$\alpha^2=1$ the algebra of constraints closes yielding the simple
result \ba
&&[C^i(p),C^j(p')]=\epsilon^{ij}\!_k D^k(p)\, \delta_{pp'}\label{eq:C1-C1}\\
&&[C^i(p), D^j(p')]=\epsilon^{ij}\!_k C^k(p) \, \delta_{pp'}\label{eq:C1-D1}\\
&&[D^i(p),D^j(p')]= \epsilon^{ij}\!_k D^k(p)\,
\delta_{pp'}\label{eq:D1-D1} \ea If we introduce the dimensionless
parameters $\gamma_0$ and $\sigma_0$ so that \[
\gamma=\sqrt{\frac{2\pi}{a_{\va H}}} \gamma_0,\ \ \ \
\sigma=\sqrt{\frac{2\pi}{a_{\va H}}} \sigma_0\] then the previous
analysis implies that we can impose spherical symmetry strongly if
and only if $(\sigma_0^2+\gamma_0^2)\pm
(\sigma_0^2-\gamma_0^2)=2$. A possible solution (for the plus
branch, or $\alpha=-1$) is for instance $\sigma_0=1$ and $\gamma_0$
arbitrary, for which the level (\ref{level}) becomes:
\be\label{kakak} k=\frac{a_{\va H}}{\frac{\kappa\beta}{4}
(1-\gamma_0^2)} \ee which exactly matches the value found in
\cite{Spherically} (see section VII) for Type I isolated horizons.
 Moreover, defining:
\be\label{Spheric} C^i_\pm(p)\equiv \frac{D^i(p)\pm
C^i(p)}{2}~~~~~~~\Rightarrow~~~~~~~ C^i_+(p)=
J_b^i(p)+J_\gamma^i(p)~~~~~~~C^i_-(p)= J_\sigma^i(p) \ee the algebra
(\ref{eq:C1-C1})--(\ref{eq:D1-D1}) becomes: \ba
&&[C^i_\pm(p),C^j_\pm(p')]=\epsilon^{ij}\!_k \ C^k_\pm(p)\, \delta_{pp'}\label{eq:Cpm-Cpm}\\
&&[C^i_\pm(p),C^j_\mp(p')]=0\label{eq:Cpm-Cmp} \ea and therefore we
can impose $C^i_\pm=0$ strongly which boils down to setting
$j^\sigma_p=0$ and $j^\gamma_p=j_p$ on the boundary spins. In this
way the Hilbert space of generic static isolated horizons
$\mathscr{H}^{CS}_k (j^\gamma_1\cdots j^\gamma_n)
\otimes\mathscr{H}^{CS}_k(j^\sigma_1\cdots j^\sigma_n)$ (restricted
only by the condition (\ref{tritri})) reduces for Type I isolated
horizons to $\mathscr{H}^{CS}(j_1\cdots  j_n) \subset {\rm
Inv}(j_1\otimes\cdots \otimes j_n)$ in complete agreement with
\cite{Spherically}. This correspondence works also at the classical
level. We have seen that spherical symmetry implies $C_{-}=0$ which,
according to Equation (\ref{eq:Classical}), requires \be
F(A_{\sigma})=0 \ee As the horizon $H$ is simply connected, this
implies that $A_\sigma=gdg^{-1}$, {\em i.e.}, pure gauge. Therefore,
the non-trivial degrees of freedom of the Type I isolated horizon
are described by a single Chern-Simons theory with connection
$A_{\gamma}$ and constraint $C_+=0$ equivalent to \be
\frac{k}{4\pi}F(A_{\gamma})=\Sigma^i \ee in complete classical
correspondence with the treatment of \cite{Spherically}. Therefore,
the requirement that spherical symmetry can be imposed strongly
reduces the two parameter family of models of a distorted horizon to
a one-parameter one. This one-parameter ambiguity is also present in
the spherical isolated horizon case (yet it does not affect the
entropy calculation).

Before ending this section let us make two important points.

The first point concerns the constraint structure at places where
$J^2_{b}=0$, {\em i.e.}, where there is no puncture. We have already
argued that at these points $J^i_{\sigma}(p)=0=J^i_{\gamma}(p)$
above, but it is instructive to briefly revisit the point here. When
$J^2_{b}(p)=0$ the constraints reduce to \ba \n &&
C^i(p)=J^i_{\sigma}(p)+J^i_{\gamma}(p)=0\\ &&
D^i(p)=J^i_{\gamma}(p)-J^i_{\sigma}(p)=0\ea with
$[C^i(p),D^j(p')]=\epsilon^{ijk} D_k(p)\, \delta_{pp'}$. There are 6
first class constraints equivalent to
$J^i_{\sigma}(p)=0=J^i_{\gamma}(p)$ and hence no degrees of freedom
at these points.

At points where $J^2_{b}(p)\not=0$ the situation is more complex. In
addition to the constraints $D^i(p)=0$ one has to impose the three
additional constraints $C^i(p)=0$ that take the form (\ref{disty})
in the arbitrarily distorted case. The problem is that, as shown
above (see Equation (\ref{eq:C-C})), the constraint algebra does not
close in the present case. Therefore, { since the set of constraints
$D^i(p), C^i(p)$ is no more first class in the generic distorted
case,} imposing the six constraints strongly is a far too strong
requirement that risks to kill relevant physical degrees of freedom.
The reason is that, in addition to the constraints $D^i(p)=0$ and
$C^i(p)=0$,  one is imposing an infinite tower of constraints
stemming from arbitrary order commutators of the original ones: we
will see that the only solutions to these are indeed spherical
configurations.

Fortunately there is a natural way of imposing the constraints
weakly. First we notice that---using the closure constraint
$D^i(p)=0${, which we do impose strongly as they are first
class---the constraint $C^i(p)$ can be written in the following form
\be\label{EPRL} ~C^i(p)\approx
J_\gamma^i(p)-J_\sigma^i(p)-\alpha(J_\sigma^i(p)+J_\gamma^i(p))=0
\ee If we interpret for a moment the previous constraint classically
we see that it implies that the vectors $J^i_{\gamma}(p)$,
$J^i_{\sigma}(p)$, and (through $D^i(p)=0$) $J^i_{b}(p)$ are
parallel. This is exactly the role of the constraint $C(p)^i=0$, as
it follows directly from the classical equations (\ref{newy}) where
one explicitly sees that all sources of curvature are proportional
to the single field $\Sigma^i$.

In order to define our strategy for a relaxation of the constraint
$C^i(p)=0$, it is particularly clarifying to take the perspective of
the master constraint technology developed by Thiemann {\em et al.}
\cite{master1,master2}. If we chose to impose the equivalent set of
(now commuting) constraints given by gauge constraint
$D^i(p)=0${---which implements the $SU(2)$ gauge symmetry at each of
the punctures---}in addition to the {\em master} constraint
$C^2(p)=0$ one would find that the only states in the kernel of
$C^i(p)$ are spherically symmetric states. Namely, the latter takes
the following explicit form:
\ba\label{master} C^2(p)=2 J_{\gamma}^2(p)+2
J_\sigma^2(p)-J_{b}^2(p)+2 \alpha (
J_{\sigma}^2(p)-J_{\gamma}^2(p))+ \alpha^2 J^2_b(p) \ea
with \be
\alpha=\frac{ \frac{a_{\va H}}{\pi} d+
(1+\gamma_0^2)}{(1-\gamma_0^2)} \ee The previous operator is
positive definite and the condition that it exists $\alpha$ such that
(\ref{master}) vanishes is given by the following restriction (stemming from the
resolvent of the previous quadratic equation) \ba\label{restriction}
&&( J_{\sigma}^2(p)-J_{\gamma}^2(p))^2- J^2_b(p) (2
J_{\gamma}^2(p)+2 J_\sigma^2(p)-J_{b}^2(p))=0\n \\ &&
J_{\sigma}^2(p)J_{\gamma}^2(p)-(J(p)_{\gamma}\cdot J(p)_{\sigma}
)^2=0 \ea where we have written it in two equivalent forms. In the
last line we see that the condition is equivalent to the vanishing
of the quantum angle $\varphi_{\sigma\gamma}$  between
$J^i_{\gamma}(p)$ and $J^i_{\sigma}(p)$, whose cosine takes the form
\be \cos(\varphi_{\gamma\sigma})=\frac{J(p)_{\gamma}\cdot
J(p)_{\sigma}}{\sqrt{J(p)_{\gamma}^2J(p)_{\sigma}^2}} \ee The only
strict  solutions of that constraint are $J^i_{\sigma}(p)=0$ or
$J^i_{\gamma}(p)=0$ which give $\alpha=\pm 1$ and hence spherically
symmetric states only. We have to relax the previous constraint and
impose it weakly. We will therefore require the previous equation to
hold only in the large spin limit. In that case it is immediate to
see that the constraint implies the simple condition \footnote{There is another branch of approximate solutions (in the large spin limit) corresponding to $j_b=|j_{\sigma}-j_{\gamma}|$; however, these correspond to antiparalel configurations which are ruled out by the form of the original constraints. See for instance (\ref{newy}).} 
\be\label{eq:weak} j_b=j^\sigma+j^\gamma \ee As shown by Major in
\cite{angle} the minimal angle goes as $1/\sqrt{j_b}$. All this
implies that it is consistent to take
  \be\label{alfa}
\alpha\equiv\frac{ J_{\sigma}^2(p)-J_{\gamma}^2(p)}{J^2_b(p)} \ee
and hence we take \be d\equiv 2\Psi_2+c=\frac{\pi}{ a_{\va
H}}\frac{(1-\gamma_0^2)(J_{\sigma}^2(p)-J_{\gamma}^2(p))-(1+\gamma_0^2)
J^2_b(p)}{J^2_b(p)} \ee in the quantum theory as well. In other
words the operator associated to the latter quantity make sense and
has no fluctuations in the Hilbert space of the distorted horizon.
Notice that Equation (\ref{alfa}) would follow from the strong
imposition of the component $C(p)\cdot J_b(p)$ of $C^i(p)$ which
together with $D^i(p)$ form a set of commuting constraints. Notice
as well that $\alpha$ defined in (\ref{alfa}) commutes with all the
observables in the boundary system! Thus, the quantity
$d=2\Psi_2+c$ remains `classical' in this sense in agreement with
the assumptions used for the construction of the phase space of our
system.

\vspace{12pt} \noindent{\bf Remark:} There is a strict analogy between the way
we have imposed the constraint $C^i(p)=0$ in this section and the
way the simplicity constraints are imposed in the EPRL-FK model
\cite{FK, EPRL}. Observe first that Equation (\ref{EPRL}) has the
very same form of the linear simplicity constraints of the EPRL-FK
models where the role of the Immirzi parameter is here played by
$\alpha$. Notice also that if we take $j_{\gamma}=(1-\alpha) j/2$
and $j_{\sigma}=(1+\alpha) j/2$ then this solves Equation
(\ref{eq:weak}) and can be checked to be consistent with alpha as
given (\ref{alfa}). With this then one can check that for an
admissible state $|\psi\rangle$ one has
$$C^2 |\psi\rangle=\hbar^2(1-\alpha^2)j |\psi\rangle$$
which vanishes in the (semiclassical) limit $\hbar\to 0$,  $j\to
\infty$ with $\hbar j$ kept constant. Moreover, using the results of
\cite{SF} one has that
$$\langle\phi |C^i|\psi\rangle=0$$
for arbitrary pairs of admissible states. In other words, the
constraint $C^i$ are satisfied strongly in the semiclassical limit,
and weakly in the sense of matrix elements in general.

\section{Entropy Calculation}\label{ec}

{ %
From now on we fix the value of $\sigma_0$ in the distorted case
according to the analysis preformed in Section \ref{2-1}, {\em
i.e.}, we set $\sigma_0=1$ and keep $\gamma_0$ as the only free
parameter }. With this preferred choice of parameters Equation
(\ref{eq:J+,J- Spherical}) now becomes: \ba\label{DJ} &&D^i(p)\equiv
J_b^i(p)+J_\gamma ^i(p)+J_\sigma ^i(p)=0 \nonumber \\ &&
J_{\gamma}^2(p)-J_\sigma^2(p)+\frac{ \frac{a_{\va H}}{\pi} (2\Psi_2+
c)+ (1+\gamma_0^2)}{(1-\gamma_0^2)} J_b^2(p)=0 \ea where, we remind,
$\Psi_2$ is the Weyl tensor component defined by $\Psi_2=C_{abcd}
\ell^a m^b \bar m^c n^d$, with $C_{abcd}$ the Weyl tensor and $\ell,
n, m$ the null tetrad defined in Statement 1 of Section \ref{ke},
which is simply related to the local scalar curvature of $H$.  The
quantity $c$ is the function relating extrinsic and intrinsic
curvature through relation (\ref{eq:KK=lambda_0 Sigma}). Both
$\Psi_2$ and $c$ are functions of the horizon points, while  Lie
dragged along the null generators $\ell^a$.

From (\ref{DJ}) we get \be d=2\Psi_2+c=\frac{\pi}{a_{\va H}}
\frac{(1-\gamma_0^2) (J^2_\sigma(p)-J_\gamma^2(p))-(1+\gamma_0^2)
J_b^2(p)}{J_b^2(p)} \ee The previous equation represents a
well-defined expression for an operator encoding the degrees of
freedom of distortion{; its eigenvalues are determined by the spins
associated to the bulk and horizon punctures and they characterize
the distorted configurations which will contribute to the entropy
calculation. More precisely, the sum over the bulk and horizon spins
performed (see below) in the states counting corresponds to the sum
over the allowed distorted configurations of the model (see remark
at the end of Section \ref{ke}). In this sense, we can trace back
the horizon entropy to the counting of the boundary geometry degrees
of freedom.}

Spherically symmetric states correspond to the eigenvalue
$-2\pi/a_{\va H}$ of the operator $2\Psi_2+c$. The deviation from
spherical symmetry can be encoded in \be \Delta\equiv
d+\frac{2\pi}{a_{\va H}}=\frac{\pi}{a_{\va H}} \frac{(1-\gamma_0^2)
(J^2_\sigma (p)-J_\gamma^2(p)+J_b^2(p))}{J_b^2(p)}
\label{eq:Delta-Psi2} \ee {in fact, the spherically symmetric
configurations correspond  to set---see Equation
(\ref{Spheric})---$j^\sigma=0$ and $j_b=j^\gamma$, giving the zero
eigenvalue of $\Delta$.}

We can now ask wether including all kind of distortions---allowing
$j^\gamma_p,j^\sigma_p$ to go all the way up to the cut off
$k/2$---preserves the area law for the IH entropy.

\subsection{The Usual Paradigm}

In this first counting we take the usual approach where the
Chern-Simons level grows with $a_{\va H}$ according to
(\ref{kakak}). Consequently we neglect the quantum corrections
coming from the (quantum group) closure constraint in consistency
with the large $k$ limit, which is equivalent to the large area
asymptotics ({\em i.e.}, the thermodynamic limit). We will show that
the inclusion of distortion does not violate the area law but simply
amounts to fixing the Immirzi parameter to value that is larger than
the one found in the spherically symmetric case.

In order to compute the number of states contained in the Hilbert
space (\ref{eq:Hilbert Space}) we are now going to use the counting
techniques of \cite{GM}. For a generic configuration of spins
associated to the punctures on the horizon we denote
$s(j^\gamma,j^\sigma)$ the occupation numbers, {\em i.e.}, the
number of punctures labeled by internal spins $j^{\gamma}$ and
$j^\sigma$ respectively. One can easily see that, keeping in mind
the constraints $j_p^{\gamma}, j_p^{\sigma}\le k/2$, for a given
configuration $\{s(j^\gamma,j^\sigma)\}$, the total number of
quantum states reads \ba\label{eq:dsj}
d({\{s(j^\gamma,j^\sigma)\}})=\frac{\left[\sum\limits_{j^\gamma,
j^\sigma}  s(j^\gamma,j^\sigma)\right] !}{\prod\limits_{j^\gamma,
j^\sigma} {s(j^{\gamma},j^\sigma)!}}\prod_{j^\gamma,
j^\sigma=0}^{\frac{k}{2}} \left( (2j^\gamma +1) (2j^\sigma
+1)\right)^{s(j^\gamma,j^\sigma)} \ea where the combinatorial factor
comes from the fact that the punctures are considered
distinguishable. To obtain the total number of states one should
then sum over all possible configuration $\{s(j^\gamma,j^\sigma)\}$.
Following \cite{GM}, we estimate the sum by maximizing $\ln
d({\{s(j^\gamma,j^\sigma)\}}) $ by varying $s(j^\gamma,j^\sigma)$
subject to the area constraint \be\label{Area constr}
\sum_{j^\gamma,j^\sigma}s(j^\gamma,j^\sigma)
\sqrt{(j^\gamma+j^\sigma)(j^\gamma+j^\sigma+1)} =\frac{a_{\va
H}}{8\pi\beta \ell_{p}^2} \ee where the above form of the area
constraint follows from the area spectrum (\ref{area}) in LQG and
the condition (\ref{eq:weak}). In the variation we assume that
$s(j^\gamma,j^\sigma) >> 1$ for each $j^\gamma,j^\sigma$ and only
such configurations dominate the counting. Introducing the Lagrange
multiplier $\lambda$, the variational equation $\delta \ln
d({\{s(j^\gamma,j^\sigma)\}})=\lambda \delta a_{\va H}$ gives
 \be\label{eq:max}
 \frac{s(j^\gamma,j^\sigma)}{\sum\limits_{j^\gamma, j^\sigma} s(j^\gamma,j^\sigma)}=(2j^\gamma +1) (2j^\sigma +1) e^{-\lambda 8\pi\beta \ell_{p}^2\sqrt{(j^\gamma+j^\sigma)(j^\gamma+j^\sigma+1)}}
 \ee
From the previous relation we obtain $\lambda\equiv
\lambda_{0}/(8\pi\beta\ell_{p}^2) $ as a solution of
\be\label{eq:lambda} 1=\sum\limits^{k/2}_{j^\gamma,j^\sigma=1/2}
(2j^\gamma +1) (2j^\sigma +1)e^{-\lambda_0
\sqrt{(j^\gamma+j^\sigma)(j^\gamma+j^\sigma+1)}} \ee obtained by
summing Equation (\ref{eq:max}) over $j^\gamma,j^\sigma$.

To estimate the leading order { in the entropy associated to the}
total number of quantum states for all configurations
$\{s(j^\gamma,j^\sigma)\}$ we evaluate the logarithm of
(\ref{eq:dsj}) at the dominant configuration (\ref{eq:max}), which
we denote $\bar{s}(j^\gamma,j^\sigma)$, namely \ba
\log{d_{\{\bar{s}(j^\gamma,j^\sigma)\}}}&\approx&
\sum_{j^\gamma,j^\sigma} s(j^\gamma,j^\sigma) \left(\log
\sum_{j^\gamma,j^\sigma} s(j^\gamma,j^\sigma)
-\log  s(j^\gamma,j^\sigma)+\log (2j^\gamma +1) (2j^\sigma +1)\right)\n\\
&=&\sum_{j^\gamma,j^\sigma}s(j^\gamma,j^\sigma)\left(\lambda
8\pi\beta
\ell_{p}^2\sqrt{(j^\gamma+j^\sigma)(j^\gamma+j^\sigma+1)}\right)
=\lambda a_{\va H} \ea where in the first line we have used
Stirling's approximation. We  conclude \be\label{ine} S\approx
\log{d_{\{\bar{s}(j^\gamma,j^\sigma)\}}} = \lambda a_{\va H}
+\sO(\log a_{\va H}) \ee where $\lambda=
\lambda_{0}/(8\pi\beta\ell_{p}^2)$ for $\lambda_0$ a solution of
Equation (\ref{eq:lambda}): numerically $\lambda_0= 2.1589...$. All
this implies that the area law is recovered for  $\beta=0.343599...$
instead of the value obtained in the spherically symmetric treatment
$\beta_{\va sph}=0.274067...$.


\subsection{A Paradigm Shift: An $a_{\va H}$-Independent  Effective Theory}\label{ps}

We have seen that, if one follows the standard paradigm, the
inclusion of distortion in the statistical ensemble produces an
entropy that grows in agreement with the area law as long as the
Immirzi parameter $\beta$ is fixed to a given numerical value. In
the present treatment though, there is an additional ambiguity
controlled by the parameter  $\gamma_0$ (as stressed at the
beginning of this section) \footnote{The presence of an analog ambiguity in the spherically symmetric case has been already emphasized in \cite{Spherically}. In the full theory such ambiguity is also present \cite{danilo} but  requirement of the introduction of a scale (such as the BH area here) at the classical level makes it less natural.}. Equation (\ref{kakak}) shows that this
ambiguity can be encoded in the value of the level $k$. Thus, the
paradigm shift that we propose consists of taking $k\in \N$ as an
arbitrary input in the construction of the effective theory
describing the phase space of generic IH{: we can make $k$ area
independent through the free parameter $\gamma_0$ by reabsorbing in
it the dependence on $a_{\va H}$. More specifically, assuming
$\gamma_0^2=(1-n\kappa\beta/(4a_{\va H}))$, where $n\in \N$, from
Equation~(\ref{kakak}) we see that the level $k$ is now free to take
any integer value $n$.}

The advantages of this is that, on the one hand, it gives a theory
which is independent of any macroscopic parameter---eliminating in
this simple way the tension present in the old treatment associated
to the natural question: {\em why should the fundamental quantum
excitations responsible for black hole entropy know about the
macroscopic area of the black hole?}---on the other hand,
compatibility with the area law will (as shown below) only fix the
relationship  between the level $k$ and the Immirzi parameter
$\beta$; thus no longer constraining the latter to a specific
numerical value.

For simplicity let us take the simplest case $k=1$; the general case
is considered at the end of the subsection. If we fix $k=1$ then the
spins at internal punctures can take only the values $j^\gamma,
j^\sigma=0,1/2$. Therefore, there are  only three possible non
trivial occupation numbers: \[A\equiv s(0,{1}/{2}), \ \ \ \ B\equiv
s({1}/{2},0),\ \ \ \ C\equiv s({1}/{2},{1}/{2})\] where the first
two cases correspond to spherically symmetric configurations.
 The total number of states for $A$, $B$, and $C$ given is
 \ba\n
 d&=&\frac{(A+B+C)!}{A! B! C!} \dim[\sH^{CS}_{1}(\underbrace{1/2\cdots1/2}_{\mbox{\small $B+C$ times}})]\dim[\sH^{CS}_{1}(\underbrace{1/2\cdots1/2}_{\va \mbox{\small $A+C$ times}})]\\
 &=&\frac{(A+B+C)!}{A! B! C!}
 \ea
since, for $k=1$, the dimension of the two $CS$ Hilbert spaces is 1.
Extremizing this number with the constraint $a_{\va H}=
[(A+B)a_{1/2}+C a_1]$ yields the following equations
 \be
 \frac{A}{A+B+C}=e^{-\lambda  a_{1/2}}, \ \ \ \
\frac{B}{A+B+C}=e^{-\lambda a_{1/2}}, \ \ \ \
\frac{C}{A+B+C}=e^{-\lambda a_{1}}
 \ee
from which it follows the condition $1=2 e^{-\lambda
a_{1/2}}+e^{-\lambda a_1}$ and a leading order entropy $S=\lambda
a_{\va H}+\sO (\log(a_{\va H}))$. All this implies that we can get
exactly Hawking area law \be S=\frac{a_{\va H}}{4\ell_p^2}+\sO
(\log(a_{\va H})) \ee if $k=1$ and the Immirzi parameter satisfies:
\be \label{esta} 1=2e^{-\pi\beta \sqrt{3} }+e^{-2\pi\beta \sqrt{2}}
\ee The previous renormalization condition is very simple and
corresponds to an area independent invariant Chern-Simons level
$k=1$. We expect a whole tower of  area independent effective
theories whose symplectic structure is given by \ba\label{boundyby}
\Omega_{\va H}(\delta_1,\delta_2)=\frac{k}{8\pi} \int_H\delta_{[1}
A_\gamma^i\wedge \delta_{2]} A_{\gamma i}- \frac{k}{8\pi}
\int_H\delta_{[1}A_{\sigma}^i\wedge \delta_{2]} A_{\sigma i} \ea
with $k\in \N$ some fixed integer, and constraints \be\label{lili}
\frac{{k}}{ 4\pi}F^i(A_{\gamma})=J_{\gamma}^i,\ \ \ \ \frac{{k}}{
4\pi}F^i(A_{\sigma})=-J_{\sigma}^i \ee with \be
J^i_{\gamma}+J^i_{\sigma}+J^i_{b}=0\ee when  $J_{b}^2 \not=0$, while
$J^i_{\gamma}=0$ and $J^i_{\sigma}=0$ when $J_{b}^2 =0$.  Provided
an appropriate modifications of the renormalization condition
(\ref{esta}) for $\beta$ {(see below)}, such effective theory for
the horizon degrees of freedom yields $S=a_{\va H}/4\ell_p^2$ to
leading order! The effective theory is independent of the
macroscopic parameter $a_{\va H}$, {\em i.e.}, the effective model
introduced here is one and the same for all IH and need not to be
tuned to the particular value of a parameter that is supposed to be
fixed only macroscopically.  For completeness we give the expression
of the operator (\ref{eq:Delta-Psi2}) controlling the deviations
from spherical symmetry which now becomes\be
\Delta=\frac{1}{8\ell_p^2\beta k} \frac{ (J^2_\sigma
(p)-J_\gamma^2(p)+J_b^2(p))}{J_b^2(p)} \ee and is independent of
macroscopic parameters.

{ Let us now consider the general case. To explicitly derive the
relation that $\beta$ has to satisfy for a generic (area
independent) value of the level $k$ such that the area law is
recovered we need to use the general formula for the dimension of
the Hilbert space $\mathscr{H}^{CS}_k (j_1\cdots j_n)$, namely
(\cite{other11,other12,other13,other21,other22,other23,other24,Terno,Karim})
\be \dim[\mathscr{H}^{CS}_k (j_1\cdots
j_n)]=\frac{2}{k+2}\sum\limits_{\ell}^{\frac{k}{2}}
\sin^2{\left(\frac{\pi (2\ell+1)}{k+2}\right)} \prod_{i=1}^{n}
\left(\frac{\sin{\left(\frac{\pi
(2\ell+1)(2j_i+1)}{k+2}\right)}}{\sin{\left(\frac{\pi
(2\ell+1)}{k+2}\right)}}\right) \ee from which we can get \be
d({\{s(j^\gamma,j^\sigma)\}})=\frac{\left[\sum_{j^\gamma, j^\sigma}
s(j^\gamma,j^\sigma)\right] !}{\prod\limits_{j^\gamma, j^\sigma}
{s(j^\gamma,j^\sigma)!}} N({\{s(j^\gamma,j^\sigma)\}}) \ee where
\ba\n
N({\{s(j^\gamma,j^\sigma)\}})&=&  \frac{4}{(k+2)^2}\sum\limits_{\ell, q=0}^{\frac{k}{2}} \sin^2{\left(\frac{\pi (2\ell+1)}{k+2}\right)} \sin^2{\left(\frac{\pi (2q+1)}{k+2}\right)}\\
\n&\cdot&\prod_{j^\gamma\ j^\sigma} \left(\frac{\sin{\left(\frac{\pi
(2\ell+1)(2j^\gamma+1)}{k+2}\right)}}{\sin{\left(\frac{\pi
(2\ell+1)}{k+2}\right)}}\right)^{s(j^\gamma)} \!\!\!\!
\left(\frac{\sin{\left(\frac{\pi
(2\ell+1)(2j^\sigma+1)}{k+2}\right)}}{\sin{\left(\frac{\pi
(2q+1)}{k+2}\right)}}\right)^{s(j^\sigma)} \ea and we have defined
$s(j^\gamma)\equiv \sum_{j^\sigma} s(j^\gamma,j^\sigma)$, and
$s(j^\sigma) \equiv \sum_{j^\gamma} s(j^\gamma,j^\sigma)$. Assuming
that \be \sum_{j^\gamma, j^\sigma=0}^{k/2} s(j^\gamma,j^\sigma)
\frac{\delta \log[ N({\{s(j^\gamma,j^\sigma)\}})] }{\delta
s(j^\gamma,j^\sigma) }-\log[
N({\{s(j^\gamma,j^\sigma)\}})]=\sO(\log(a_{\va H})) \ee we obtain
that $\beta$ has to satisfy \be\label{esta2} 1=\sum_{j^\gamma,
j^\sigma=0}^{k/2}\exp {\left(\frac{1}{ N({\{s(j^\gamma,j^\sigma)\}})
}\frac{\delta N({\{s(j^\gamma,j^\sigma)\}})}{\delta s(j^\gamma
,j^\sigma)}\right)}\exp{\left( -2\pi\beta
\sqrt{(j^\gamma+j^\sigma)(j^\gamma+j^\sigma+1)} \right)} \ee The
previous expression encodes the relationship between $k$ and
$\beta_k$ dictated by the validity of the area law. Analytic
information could be extracted from it by the usual approximation
methods or it could be used as the basis for numerical computations.
Some exact as well as qualitative information in shown in Figure
\ref{gura}.

\begin{figure}[H]
\centerline{\hspace{0.6cm} \(
\begin{array}{c}
\includegraphics[height=6cm]{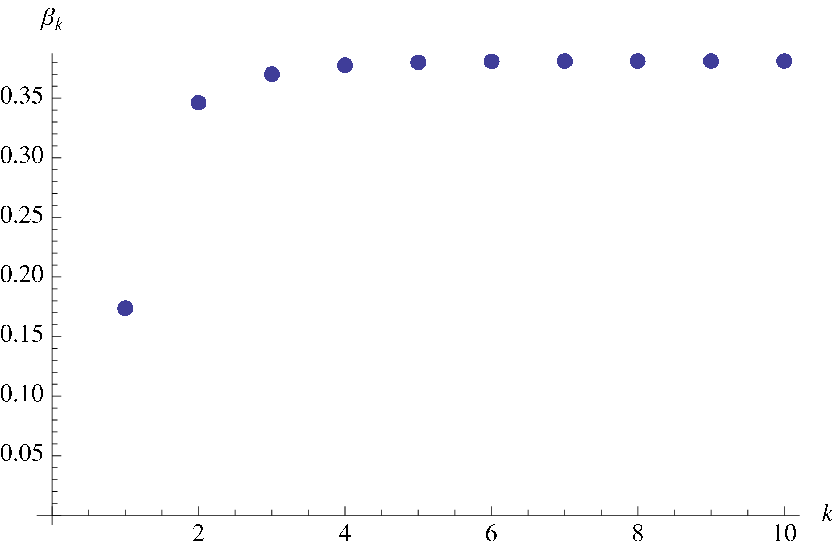}
\end{array}
\label{beta}\)} \caption{The value of the Immirzi parameter
$\beta_{k}$ as a function of $k\in \N$ for the first few integers.
The value $\beta_{1}=0.172217...$ is exact as well as the asymptotic
value $\beta_{\infty}=0.343599...$. The other points have been
computed using (\ref{eq:lambda}) which is only valid in the large
$k$ limit. } \label{gura}
\end{figure}

\section{Conclusions}\label{conclusion}

The main results of this paper can be organized in three parts. In
the first part we have shown that it is possible to describe the
classical phase space of distorted isolated horizons without the
need to invoke the notion of any Type I structure as it was
generally done in previous work \cite{AEV, jony}. Our treatment does
not require any symmetry assumption treating (static) distorted
isolated horizons of type I, II or III on equal footing. Non-static
horizons are more complicated systems as our analysis shows due to
the breaking of diffeomorphism invariance at the boundary. As we
explain in Section \ref{ns}  the breaking of diffeomorphism
invariance is only a problem from the point of view of the
quantization in the context of loop quantum gravity, as this gauge
symmetry breaking is in manifest contradiction to the gauge
invariance of the bulk theory. We  propose a way to restore
diffeomorphism invariance by the enlargement of the kinematical
phase space. This strategy is certainly viable, however a clear-cut
interpretation of the new system as well as its quantization
requires further study. It is important to point out that, as the
stationary Kerr-Newman black hole is a non-static isolated horizon,
further insights on the issues here discussed are necessary for a
complete understanding of the nature of the black hole entropy
calculation including rotating horizons.

In the case of static isolated horizons we show that the classical
system can be described in terms of certain connection variables. We
show that there is a two-parameter family of such variables in terms
of which the pre-symplectic structure of the system acquires a
boundary term corresponding to the sum of two (closely related)
$SU(2)$ Chern-Simons theories. In addition to the pre-symplectic
form with its boundary terms, there are constraints that relate the
bulk and boundary fields at the isolated horizon and ensure the
validity of the boundary condition. The free parameters describing
the classically equivalent formulations are the analog of the
Immirzi parameter in the full theory in the sense that, as in the
latter case, they appear when one changes from the one-form variable
phase space parametrization---see Equation (\ref{bb})---to the
connection parametrization---see Equation (\ref{boundy})---that
makes the quantization in loop variables more natural. Remarkably,
the ambiguity can be entirely encoded in the value of the level of
the Chern-Simons theories $k\in \N$:  there is an integer worthy of
theories as it becomes clear when rewriting (\ref{boundy}) as in
Equations (\ref{boundyby}) with sources as in (\ref{lili}). The
conservation of the presymplectic structure proven in Section
\ref{sec:conserved symplectic structure} goes though without making
reference to $a_{\va H}$ and $\gamma_0$.

In the second part of the paper we study the quantization of the
static isolated horizon phase space. The quantization can be defined
along  similar lines as the one followed in the spherically
symmetric case. However, subtleties arise in the imposition of the
quantum boundary conditions. In particular we observe in Section
\ref{2-1} that the (boundary) constraint algebra does not close in
general. We use this fact to reduce the two-parameter family of
classically equivalent formulations to a one-parameter family by
requiring that the constraint algebra closes in the case of Type I
({\em i.e.}, spherically symmetric) isolated horizons. Remarkably,
this yields a quantization that is precisely equivalent to the one
presented in \cite{prl,Spherically}. For distorted isolated horizons
this implies that one can impose all the boundary constraints only
weakly: there is a natural way to impose them in the large spin
limit by restricting to those configurations that minimize the
`angle' between the Chern-Simons defects.

We compute the entropy from the quantum theory by counting states in
an ensemble containing distorted horizons of a given macroscopic
area $a_{\va H}$ and observe that the inclusion of distortion (in
the standard treatment) does not violate the area law, but simply
slightly increases the numerical value the Immirzi parameter has to
be fixed to. {In this regard, it is interesting to notice that this
regulating effect of the quantum group which somehow encodes the
distortion structure is a consequence of the proper way we have
imposed the constraints. In fact, without the restriction
(\ref{eq:weak}), {\em i.e.}, allowing every angle between the two
sets of punctures on the horizon, it can be shown that the entropy
would grow as $a_{\va H} \log (a_{\va H})$ in violation of the area
law.}

In the third part (Section \ref{ps})  we show that the $SU(2)$
treatment allows for a simple paradigm shift where instead of tuning
the Immirzi parameter in order to get agreement with Hawkings
formula---the usual approach followed in the standard
literature---we take the level of the Chern-Simons theory as an
independent effective parameter (independent of the area $a_{\va
H}$). The advantages of such treatment are multiple. On the one hand
the effective theory so defined is universal: independent of any
macroscopic parameter such as the area $a_{\va H}$, {\em i.e.}, the
same for all isolated horizons. On the other hand,  from this new
perspective one gets agreement with Hawking  area law without the
need of fixing the Immirzi parameter to  a single particular value
as there is a integers-worth of possible values of $\beta_{k}$
satisfying the relation (\ref{esta2}) for  $k\in \N$. From this
viewpoint the semiclassical consistency with the Hawking effect is
not viewed as a restriction on a fundamental constant in LQG (the
Immirzi parameter) but rather as providing insights of a more
fundamental (dynamical) description underlying the necessary
relationship---manifested here by the admissible effective
descriptions labeled by $(k, \beta_k)$ \footnote{Notice, that the family of descriptions here presented are labelled
by an integer $k\in \N$ and a correlated real number. This seem to
be the structure that one would wish in order to establish a link
between LQG and some interesting speculative ideas explored recently
\cite{hanhan}.}---between
the LQG in the bulk and the $SU(2)$ Chern-Simons theory on the
boundary. None of two characteristic constants of the bulk theory
(Immirzi parameter) and of the boundary theory (CS level) is more
important than the other in this model: only a given relationship
among the two is required in order to recover the Bekenstein-Hawking
formula for black holes entropy \footnote{It interesting to speculate a possible understanding of our result
from the point of view of renormalization, as emphasized by Jacobson
in \cite{jaco}.}.  The range of
physically suitable values of $\beta_{k}$ is illustrated in Figure
\ref{gura}.

Moreover, the model presented here allows for counting states
corresponding to generic isolated horizons,
eliminating in this way a puzzling physical restriction imposed in
the entropy count of Type~I~horizons.

{It is important to stress that the possibility of taking the
Chern-Simons level as an independent parameter is a feature of the
$SU(2)$ analysis due to the appearance of a one-parameter freedom
lacking in the $U(1)$ case. In particular, the view proposed in
Section \ref{ps} could have been taken as well in the definition of
the Type I effective theory leading to the entropy calculation of
\cite{Spherically}}. If such would have been the view then, only a
relationship between the Immirzi parameter and the Chern-Simons
level would have been required in order to satisfy the area law in
this symmetry reduced case as well.

\newpage
\section*{Acknowledgements}

We thank Eugenio Bianchi, Jonathan Engle,  Laurent Freidel, Karim
Noui  for discussions. We also thank the various remarks and
suggestions that we got from the three anonymous referees. This work
was supported in part by the Agence Nationale de la Recherche; grant
ANR-06-BLAN-0050. A.P. was supported by {\em l'Institut
Universitaire de France}. D.P. was supported by {\em Marie Curie}
EU-NCG network.






\bibliographystyle{mdpi}

\makeatletter

\renewcommand\@biblabel[1]{#1. }

\makeatother





\end{document}